\documentclass[twocolumn]{aastex631}

\usepackage{amsmath}
\usepackage{xcolor}

\newcommand{\be}{\begin{equation}}
\newcommand{\ee}{\end{equation}}
\newcommand{\bea}{\setlength\arraycolsep{2pt} \begin{eqnarray}}
\newcommand{\eea}{\end{eqnarray}}
\newcommand{\nn}{\nonumber}
\newcommand{\mA}{{\mathcal A}}
\newcommand{\mB}{{\mathcal B}}
\newcommand{\mO}{{\mathcal O}}
\newcommand{\mE}{{\mathcal E}}

\newcommand{\mL}{{\mathcal L}}

\newcommand{\mQ}{{\mathcal Q}}

\def\a{\alpha}
\def\b{\beta}
\def\c{\chi}
\def\d{\delta}
\def\D{\Delta}
\def\f{\frac}
\def\g{\gamma}
\def\G{\Gamma}
\def\lm{\lambda}
\def\l{\left}
\def\L{\Lambda}
\def\m{\mu}
\def\n{\nu}
\def\pl{\partial}
\def\p{\phi}

\def\r{\right}
\def\s{\sigma}
\def\S{\Sigma}
\def\t{\theta}
\def\T{\Theta}
\def\k{\kappa}

\def\o{\omega}

\begin{document}

\title{Near-Horizon Polarization as a Diagnostic of Black Hole Spacetime}

\correspondingauthor{Minyong Guo}
\email{minyongguo@bnu.edu.cn}
\correspondingauthor{Yosuke Mizuno}
\email{mizuno@sjtu.edu.cn}
\correspondingauthor{Bin Chen}
\email{chenbin1@nbu.edu.cn}

\author[0000-0002-9434-3930]{Yehui Hou}
\altaffiliation{These authors contributed equally to this work.}
\affiliation{Tsung-Dao Lee Institute, Shanghai Jiao-Tong University, Shanghai, 201210, P. R. China}

\author[0009-0002-2360-2971]{Jiewei Huang}
\altaffiliation{These authors contributed equally to this work.}
\affiliation{School of Physics, Peking University, No.5 Yiheyuan Rd, Beijing 100871, P.R. China}

\author[0000-0001-5577-575X]{Minyong Guo}
\affiliation{School of Physics and Astronomy, Beijing Normal University, Beijing 100875, P. R. China}
\affiliation{Key Laboratory of Multiscale Spin Physics, Ministry of Education, Beijing 100875, P. R. China}

\author[0000-0002-8131-6730]{Yosuke Mizuno}
\affiliation{Tsung-Dao Lee Institute, Shanghai Jiao-Tong University, Shanghai, 201210, P. R. China}
\affiliation{School of Physics \& Astronomy, Shanghai Jiao-Tong University, Shanghai, 200240, P. R. China}
\affiliation{Key Laboratory for Particle Physics, Astrophysics and Cosmology (MOE), Shanghai Key Laboratory for Particle Physics and Cosmology, Shanghai Jiao-Tong University, Shanghai, 200240, P. R. China}

\author[0000-0003-4509-9705]{Bin Chen}
\affiliation{Institute of Fundamental Physics and Quantum Technology, Ningbo University, Ningbo, Zhejiang 315211, China}
\affiliation{School of Physical Science and Technology, Ningbo University, Ningbo, Zhejiang 315211, China}
\affiliation{School of Physics, Peking University, No.5 Yiheyuan Rd, Beijing 100871, P.R. China}
\affiliation{Center for High Energy Physics, Peking University, No.5 Yiheyuan Rd, Beijing 100871, P. R. China}

\begin{abstract}
A key challenge in imaging supermassive black holes is disentangling gravitational effects from plasma physics in order to accurately determine spacetime properties, particularly black hole spin. 
In this Letter, we present a fully covariant and rigorous analysis of the synchrotron emission from  accreting plasma in the equatorial plane in the stationary, axisymmetric, high-conductivity regime, and identify--for the first time--a distinctive near-horizon polarization pattern that remains robust across different flow structures.
This pattern arises from strong frame dragging near the event horizon, which induces a degeneracy among plasma flow and magnetic field configurations, yielding a polarization signature determined solely by the spacetime geometry and the observer's inclination. The near-horizon polarization thus offers a clean and precise probe of black hole spin and other fundamental parameters. 
If future space-based millimeter VLBI observations can resolve synchrotron emission originating within approximately 1$\%$ of the event horizon radius in M87* or Sgr A*, this universal polarization pattern may become observable.
\end{abstract}


\section{Introduction}
The supermassive black hole images obtained by the Event Horizon Telescope (EHT) Collaboration have opened an unprecedented window for exploring strong gravitational fields \citep{EventHorizonTelescope:2019dse, EventHorizonTelescope:2019pgp, EventHorizonTelescope:2019ggy, EventHorizonTelescope:2021bee, EventHorizonTelescope:2021srq, EventHorizonTelescope:2022wkp, EventHorizonTelescope:2022urf, EHT:2024nwx, EHT:2024ehx}.
The strong-field gravity of black holes leaves distinctive imprints on these images \citep{Gralla:2020srx, Medeiros:2021apx, Dokuchaev:2019jqq, Hadar:2020fda}, enabling the investigation of extreme astrophysical environments and allowing constraints to be placed on the black hole's mass, spin, and potential deviations from the Kerr metric based on observational data.
In particular, polarimetric observations provide an especially sensitive probe of the horizon-scale environment \citep{Moscibrodzka:2017gdx, Ricarte:2020llx, Emami:2022kci}. 
The polarization vector of synchrotron radiation is intrinsically orthogonal to the local magnetic field in the plasma comoving frame \citep{1979Lightman}. However, this orthogonality is modified by the rapid accretion flow of the plasma and by gravitational lensing \citep{Gelles:2021kti}, such that the observed polarization pattern encodes information about the magnetic field structure, accretion dynamics, and the underlying spacetime geometry.

Magnetized plasmas exhibiting different inflow modes produce distinct magnetic field configurations, which in turn significantly modify the observed polarization morphology \citep{Vincent:2020dij, Ricarte:2022wpd, Chael:2023pwp}. By comparing the synthetic images of accretion flows generated from general relativistic magnetohydrodynamics (GRMHD) simulations with observational data, it is possible to qualitatively constrain the underlying magnetic field structure and discriminate between two extreme black hole spin scenarios—rapidly spinning and slowly spinning black holes \citep{Palumbo:2020flt, Qiu:2022fzl, Zhang:2024owe, Palumbo:2024jtz}. Nevertheless, the finite angular resolution of telescopes and the complex origins of polarization patterns make it challenging to accurately disentangle gravitational effects from astrophysical ones, limiting our ability to reliably quantify the role of spacetime geometry. Therefore, identifying polarization observables dominated by gravity and insensitive to plasma physics would be highly valuable for both theory and observation.

In this Letter, we show that near-horizon physics offers a promising path forward. In the vicinity of rotating black holes, frame dragging twists both inflowing plasma and magnetic fields into spiral structures \citep{Ricarte:2022wpd}. Sufficiently close to the horizon, this effect drives a variety of inflow modes toward a predominantly toroidal configuration. We anticipate that such strong frame dragging induces a universal near-horizon polarization (NHP) pattern on the image plane, largely independent of the details of accretion. For sources such as M87* and Sgr A*, the inner accretion flow likely resides in a magnetically arrested disk (MAD) state \citep{Narayan:2003by, Tchekhovskoy:2011zx, EventHorizonTelescope:2021srq}, with emissivity sharply rising toward the horizon \citep{Chael:2018gzl, Narayan:2021qfw, Chael:2021rjo}, enabling photons to escape despite strong gravitational redshift. This makes the NHP pattern potentially observable with future EHT arrays, especially via space-based VLBI \citep{Lockhart:2022rui, Ayzenberg:2023hfw, Johnson:2023ynn, Jia:2024mlb, Johnson:2024ttr, EventHorizonTelescope:2024whi}.

Under the assumptions of stationarity and axisymmetry, we perform a fully covariant analysis of equatorial plasma flows characterized by ideal conductivity and negligible vertical magnetic field component, along with the associated polarized synchrotron emission profile. For the Kerr spacetime, by expanding the flow velocity, magnetic field, and polarized emission in powers of the radial distance from the horizon, and employing symmetry-induced lensing and polarization transfer, we derive a general NHP formula. Remarkably, the leading-order NHP pattern depends solely on the spacetime geometry, while the next-to-leading term is influenced by both the spacetime and the field line angular velocity, but remains insensitive to the inflow velocity.

We present the analytic NHP pattern on the observer's image plane and demonstrate its strong dependence on black hole spin. Especially, for a nearly on-axis observer, the secondary azimuthal Fourier mode $\b_2$ of linear polarization \citep{Palumbo:2020flt} takes an elegant, model-independent form at leading order:
\bea\label{NHPbeta}
\b_2 \, \approx \, -2\tan^{-1} \left[ \f{a}{\sqrt {4r_+-a^2}} \,\,\right]
\eea
where $a$ denotes the black hole spin parameter and $r_+$ is the horizon radius. 
Our analytic NHP pattern enables direct, non-degenerate spin measurements. 
Furthermore, we verify that the NHP formula can be directly extended to the Kerr-Newman-Taub-NUT spacetime, suggesting its universality for a broad class of rotating black holes.
Throughout this work, we adopt natural units with $GM = c = 1$, where $M$ denotes the black hole mass.

\section{The magnetized accretion flow}
 
We begin with a steady-state accretion of a magnetized plasma onto a general stationary, axisymmetric black hole, assuming that the plasma's self-gravity is negligible. 
The plasma is taken to be perfectly conducting, so that the electric field vanishes in the comoving frame, i.e., $e^{\m} = F^{\m\n}u_{\n} = 0$, where $F^{\m\n}$ is the electromagnetic field tensor and $u^{\m}$ is the plasma four-velocity. 

In a stationary and axisymmetric configuration, the electromagnetic field is governed by two conserved quantities along each magnetic field line (or equivalently, each streamline) 
\citep{thorne1982electrodynamics, 1984PhDT........92P}: the stream function $\psi$, which labels magnetic flux surfaces, and the field line angular velocity $\Omega_B$, which characterizes the rotation of magnetic field lines. 
In terms of spherical coordinates $\{t,r,\t,\p\}$, the magnetic field as measured in the comoving frame can be expressed as
\bea
b^{\m} &=& -(*F)^{\m\n}u_{\n}  \nn \\
&=&  -\f{\pl_{\t}\psi}{\sqrt{-g}\,u^r} \left[ \left( u_t + \Omega_B u_{\p} \right)u^{\m} + \d^{\m}_t + \Omega_B\d^{\m}_{\p} \right] \, , 
\label{bmu}
\eea
where $g$ is the determinant of the metric tensor.

The field line angular velocity $\Omega_B$ induces a nonzero electric field in the black hole coordinate frame, enabling energy extraction from black hole rotation via relativistic jets. The extraction efficiency for a Kerr black hole is maximized for $\Omega_B \approx \Omega_H/2$ within the jet region \citep{Blandford:1977ds}. Numerical simulations further indicate that $\Omega_B \approx 0.2\,\Omega_H - 0.5\,\Omega_H$ within MADs \citep{Tchekhovskoy:2011zx, mckinney2012general}.

In the MAD regime, the accumulation of magnetic flux near the event horizon generates a strong magnetic pressure that compresses the accretion flow toward the equatorial plane \citep{Chael:2018gzl, Chael:2021rjo, Ricarte:2022wpd}. In this region, the flow becomes predominantly radial, satisfying $|r u^{\t}| \ll |u^r|$. The conservation of the stream function simplifies to $u^{\m} \pl_{\m} \psi \approx  u^r \pl_r \psi \approx 0$, implying that $\Psi \equiv \pl_{\t}\psi$ can be treated as approximately constant near the equatorial plane \citep{Hou:2023bep} (see Appendix.~\ref{Appmag} for details). 
Under these conditions, Eq.~\eqref{bmu} reveals a predominantly radial magnetic field structure, with field lines being advected inward by the plasma flow. This behavior is supported by both GRMHD simulations \citep{Komissarov:2005wj, Tchekhovskoy:2011zx, mckinney2012general} and particle-in-cell simulations \citep{Parfrey:2018dnc}. We do not consider purely circular flows with $u^r = 0$, as they are disfavored by current polarimetric observations \citep{Palumbo:2024czv}.

Notably, in parabolic or more collimated jets extending away from the equatorial plane, the flow velocity and magnetic field tend to develop significant vertical components \citep{Tchekhovskoy:2011zx}. Nevertheless, the jet emission is typically subdominant relative to that of the accretion flow. As such, we neglect the jet's emission profile in the present study.

\section{Polarized emission profile}

At millimeter wavelengths, the emission is dominated by (thermal or non-thermal) synchrotron radiation from electrons within the plasma, whose unit-norm polarization vector $f^{\m}$ is largely perpendicular to the magnetic field \citep{1979Lightman}. Thus, without specifying a particular electron distribution, the form of $f^{\m}$ can be determined by geometric constraints:
\bea\label{fmu}
f^\m_{(e)} = \f{\epsilon^{\m\n\rho\sigma} u_\n p_\rho b_{\sigma} }{\omega\sqrt{b_{\perp}^2} }\, ,
\eea
where $\epsilon^{\m\n\rho\sigma}$ is the Levi-Civita tensor, $p^{\m}$ is the photon four-momentum, $\omega = -u^{\m} p_{\m}$ is the photon frequency measured in the comoving frame. We have introduced the spatial magnetic field in the comoving frame $b^{\m}_{\perp} = b^\m - \omega^{-2}(b\cdot  p) \, p_{\perp}^\m$ to properly normalize $f^{\m}$, where $p_{\perp}^\m = p^\m -\omega u^\m$ is the spatial component of the photon four-momentum. Here, the subscript ``$(e)$'' denotes the emission location.

After emission, photons propagate along null geodesics, with their polarization vectors parallel transported along the trajectories. The polarization condition and transport equations are given by $f^{\m}p_{\m}=0$, $p^{\m} \nabla_{\m} p^\n = p^{\m} \nabla_{\m} f^\n = 0$. The polarization vector admits a gauge freedom $f^\m \rightarrow f^\m + \xi \, p^\m$, where $\xi$ is an arbitrary constant. 

The arrival positions of photons on the image plane of a distant, static observer are described by the image coordinates: $(\alpha, \beta) \propto (-\sin{\theta_o} p^{\phi}, p^{\theta})$, where $\theta_o$ is the observer's inclination angle. The image plane is thus spanned by the tetrad $\{ e^{[\a]}_{\m},e^{[\b]}_{\m} \}$ \citep{Bardeen:1973tla}. The observed polarization pattern can be characterized by the electric vector polarization angle (EVPA) of the arriving photons, computed from the components of $f^{\m}$ projected along the $\a$ and $\b$ directions \cite{EventHorizonTelescope:2021srq}:
\bea\label{eevpa}
&&\chi =  \tan^{-1}{ \left[ -\f{f_{(o)}^{\m}e^{[\a]}_{\m}}{f_{(o)}^{\m}e^{[\b]}_{\m}} \right] }\,,
\eea
which measures the polarization direction relative to the $\b$-axis. Here, the subscript ``$(o)$'' denotes the observer's location.

\section{Near-horizon limit}

In this section, we focus on the Kerr geometry to analytically investigate polarization behavior in the vicinity of the event horizon. In Boyer-Lindquist coordinates, the event horizon is defined by the vanishing of the redshift factor,
\bea
\D \equiv r^2 -2r+a^2 = (r-r_-)(r-r_+) = 0 \,,
\eea
where $a$ is the black hole spin parameter and $r_{\pm} = 1 \pm \sqrt{1-a^2}$. The physical event horizon is located at $r = r_+$. Near the horizon, $\D \approx \left( r_+ - r_- \right)\d r$, with $\d r = r - r_+$, allowing the use of the redshift factor $\D$ as a measure of the radial distance to the horizon.

In the near-horizon limit $\D \rightarrow 0$, all relevant quantities are expanded as power series in $\D$ on the equatorial plane. Without loss of generality, we set $u^\t=0$ and hold $u_t$, $u_\p$ constant during the expansion. The components of the flow velocity are then given by
\bea\label{eou}
\begin{aligned}
&u^t = -\f{4Y}{\D} - \left(1+\f{2}{r_+}\right)u_t+\f{a u_{\p}}{r_+^2(r_+-1)} + \mO(\D) \, ,  \\
&u^r=  -\f{2|Y|}{r_+} + \mO(\D) \, , \,\, u^{\t} = 0 \, ,  \\
&u^{\p} =  -\f{2aY}{r_+\D} + \f{au_t+ u_{\p}}{r_+^2(r_+-1)} + \mO(\D) \, ,
\end{aligned}
\eea
where $Y=u_t+\Omega_H u_{\p}$, $\Omega_H = a/(2r_+)$ is the angular velocity of the Kerr horizon. The radial velocity $u^r$ determined by the normalization $u^{\m}u_{\m} = -1$, and we choose $u^r < 0$ to describe accreting plasma. To leading order, $u^{\p}/u^{t} \approx \Omega_H$ and $u^r/u^t \approx \sigma_Y a^{-1}\Omega_H\D$, where $\sigma_Y$ denotes the sign of $Y$. 
For plunging matter, $\sigma_Y < 0$, in consistent with the second law of black hole thermodynamics \citep{Chandrasekhar:1985kt}; specifically, a test particle with $Y > 0$ cannot classically reach the horizon. 
In Eq.~\eqref{eou}, $u_t$ and $u_{\p}$ are two free functions of $r$ outside the horizon. They constitute the two degrees of freedom of equatorial plasma motion and determine the specific inflow structure.

The near-horizon expansion of the magnetic field, as given by Eq.~\eqref{bmu}, reads:
\bea\label{eob}
\begin{aligned}
&b^t = \f{2 \Psi }{r_+\D} \left( u_t + \Omega_B u_{\p} \right) + \mO(1) \, ,  \\
&b^r = -\f{\Psi}{r_+^2} \left( u_t + \Omega_B u_{\p} \right) \, , \, \, b^{\t} = 0 \, ,  \\
&b^{\p} = \f{ \Psi a}{r_+^2 \D} \left( u_t + \Omega_B u_{\p} \right) + \mO(1) \, .
\end{aligned}
\eea
The ratio $b^{\phi}/b^r \approx -a \, \Delta^{-1}$ reveals a highly toroidal magnetic field structure near the horizon \citep{Ricarte:2022wpd}.
The vanishing of $b^{\t}$ arises from the assumption that $u^{\theta} = 0$ -- since the vertical velocity is rather small compared to the radial one near the equatorial plane, according to Eq.~\eqref{bmu}, we have $r|b^{\t}| \ll |b^{r}|$.

In the Kerr spacetime, photon trajectories are characterized by three constants of motion: the energy $\mE = -p_t$, angular momentum $\mL = p_{\p}$, and the Carter constant $\mQ = p_{\t}^2 + a^2\mE^2\cos^2{\t} + \mL^2 \cot^2{\t}$. The $r,\t$ components of the photon momentum are determined by effective potentials \citep{Carter:1968rr}. 
By expanding the photon four-momentum and employing Eqs.~\eqref{eou} and \eqref{eob}, the near-horizon expansion for the polarization vector at the emission point is
\bea\label{eof}
\begin{aligned}
X f^{t}_{(e)} = \,\, &\f{4\sigma_{\t}\sqrt{\eta}}{ \D }\left( \Omega_B - \Omega_H \right) + \f{\sigma_{\t}\sqrt{\eta}}{r_+}(r_++2)\Omega_B   \\
& + \f{a\sigma_{\t}\sqrt{\eta}}{r_+^2(r_+-1)} + \mO(\D) \,,\\
X f^{r}_{(e)} =  \,\, &\f{2\sigma_{\t}\sqrt{\eta}}{r_+} \left( \Omega_B - \Omega_H \right)  + F_r \D + \mO(\D) \,,  \\
X f^{\t}_{(e)} = \,\, &- 4 \f{(\Omega_H\lm-1)-\sigma_r |\Omega_H\lm-1|}{\D}\,\left( \Omega_B - \Omega_H \right)   \\
&+ F_{\t}+ \mO(\D)  \,, \\
X f^{\p}_{(e)} =  \,\,&\f{2a\sigma_{\t} \sqrt{\eta}}{ r_+ \D }\left( \Omega_B - \Omega_H \right)  \\
& + \f{\sigma_{\t} \sqrt{\eta}}{ r_+^2(r_+-1)}\left( 1- a\Omega_B \right) + \mO(\D) \,,
\end{aligned}
\eea
where $X = -\Psi^{-1} \mE^{-1}\omega \sqrt{-g \, b_{\perp}^2}\,$; $\sigma_r,\sigma_{\t}$ denote the signs of $p_r$ and $p_{\t}\,$; $\lm \equiv \mL/\mE, \eta \equiv \mQ/\mE^2$ are the photon impact parameters, which specify the emission direction \citep{cunningham1973optical}. We restrict to $\eta > 0$, corresponding to light rays that can reach the equatorial plane.

The leading-order terms in Eq.\eqref{eof} are all proportional to $\left( \Omega_B - \Omega_H \right)$, and do not explicitly depend on $u^{\m}$. Nevertheless, since the prefactor $X$ depends on $u^{\m}$, the leading-order polarization vector retains an implicit dependence on the flow velocity. The terms $F_r$ and $F_{\t}$ in Eq.\eqref{eof} are functions of the flow velocity, but their explicit forms are lengthy and are omitted here for brevity.

\section{The NHP formula}

Since the Kerr spacetime is of Petrov Type D \citep{petrov1954classification}, there exists a conserved quantity associated with the parallel transport of the polarization vector along a null geodesic: the complex Penrose-Walker constant \citep{Walker:1970un},
\bea \label{PW11}
\begin{aligned}
&\kappa \equiv \k_{1}+i \k_{2} = (r-ia\cos\t)(\mA-i\mB) \, , \\
&\mA= 2p^{[t} f^{r]} + 2a \sin^2{\t} p^{[r} f^{\p]} \, ,  \\
&\mB = 2\sin{\t}\left[(r^2+a^2) p^{[\p} f^{\t]} -a p^{[t} f^{\t]}\right] \,,
\end{aligned}
\eea
where $\kappa_1, \kappa_2$ denote the real and imaginary parts of $\kappa$, respectively. According to Eq.~\eqref{PW11}, the parallel transport of the polarization vector reduces to an algebraic problem: its two degrees of freedom, $\{ f^{\a}_{(o)},f^{\b}_{(o)} \}$, can be expressed in terms of $\kappa_1$ and $\kappa_2$, which are determined by $f^{\m}_{(e)}$ at the emission point.

Using the near-horizon expansion for $f^{\m}_{(e)}$ (Eq.~\eqref{eof}) and the expansion of the photon four-momentum, we find that the ratio
\bea\label{master}
z \equiv \f{\kappa_1}{\kappa_2} = z_0 + z_1 \Delta +\mO(\Delta^2)
\eea
is independent of the flow velocity up to next-to-leading order. Here, we introduce the following expressions:
\bea\label{z0z1}
\begin{aligned}
& z_0 = \f{\sigma_\t \sqrt{\eta}}{\lm - a}\,, \\
& z_1 = \Omega_H^2 \, \f{(1-a\Omega_B)}{(\Omega_H -\Omega_B)} \, \f{\sigma_\t \left(1 + z_0^2\right) \sqrt{\eta}}{2a^2 (\Omega_H \lm - 1)}\,.
\end{aligned}
\eea
It should be noted that Eq.\eqref{master} is valid below the superradiant bound, i.e., $\lm \leq 1/\Omega_H$, which ensures that photons from near-horizon emission with positive energy (and thus observational relevance) satisfy this condition \citep{Compere:2021bkk}.

We then propagate the near-horizon expansion for $f^{\m}_{(e)}$ to the observer's image plane to obtain the NHP pattern. In the Kerr spacetime, the image coordinates of the arriving photons are related to the impact parameters \citep{Bardeen:1973tla}: $\a = -\lm (\sin{\t_o})^{-1}\,$, $\b^2 = \eta +(a^2 -\a^2) \cos^2{\t_o}\,$. From Eq.\eqref{eevpa} and Eq.\eqref{PW11}, the distribution of the EVPA (Eq.\eqref{eevpa}) on the image plane takes the form of \cite{Himwich:2020msm}
\bea\label{EVPAkerr}
\chi = \tan^{-1}\left(\frac{\mu z- \beta }{\beta z + \mu}\right)
\,,\quad \m = -(\a+a \sin{\t_o}) \,.
\eea
Combining Eqs.\eqref{master} and \eqref{EVPAkerr}, we obtain the near-horizon EVPA as
\bea\label{efin}
\chi =  -\tan^{-1}\left(\frac{\beta}{\mu}\right) + \tan^{-1}z_0 + \frac{z_1 \Delta}{1 + z_0^2} + \mO(\Delta^2)\,.
\eea
In this expansion for $\chi$, the leading-order term depends only on the black hole spin parameter and the image coordinates $(\a,\b)$. \footnote{It is well established that the first-order geodesic equations in the $r$ and $\t$ directions in the Kerr spacetime allow the emission-image mapping to be expressed via geodesic integrals, yielding $\D = \D(\a,\b,a,\t_o)$ \citep{Beckwith:2004ae,Gralla:2019drh}.} The next-to-leading order term is determined by the spin parameter, image coordinates, and the field line angular velocity $\Omega_B$. 
Therefore, the near-horizon EVPA distribution on the image plane Eq.~\eqref{efin} is independent of the flow structure, being determined solely by the spacetime geometry and the magnetic field rotation. In particular, this result highlights a unique imprint of the black hole spin on the NHP.
Note that when using Eq.~\eqref{efin}, a modulus should be applied to $\chi$ to confine it to $[-\pi/2, \pi/2]$, since the arctangent sum may exceed this range.

To assess the convergence radius of the near-horizon expansion, we compare $z_0(a,\lm,\eta)+ z_1(a,\lm,\eta,\Omega_B)\D$ from Eq.\eqref{master} with the exact value $z(a,\D,\lm,\eta,\Omega_B,u^{\m})$ for different inflow models characterized by distinct $u_t$ and $u_\p$: free-fall (Type I), plunging from the prograde innermost stable circular orbit (ISCO) (Type II), and the fit from the MAD simulation (Type III) \citep{Chael:2021rjo}. The comparison is shown in Fig.\ref{test}, where representative values of $\lm,\eta$ and $\Omega_B$ are chosen (see Appendix.~\ref{appcompare} for results for a broader parameter space). 
As shown in Fig.\ref{test}, the expansion in Eq.\eqref{master} up to next-to-leading order in the redshift factor is accurate for $ 0 \leq \Delta \lesssim 10^{-2} $ (corresponding to $ r = 1.01r_+ $ for $ a = 0.94 $ and $ r = 1.003r_+ $ for $ a = 0.2 $) across all three flow models. Inclusion of the next-to-next-to-leading order term further improves the accuracy only in a very narrow region around $ \Delta \simeq 5 \times 10^{-2} $ (corresponding to $ r = 1.05r_+ $ for $ a = 0.94 $ and $ r = 1.01r_+ $ for $ a = 0.2 $). 

\begin{figure}[ht!]
	\centering
	\includegraphics[width=3.2in]{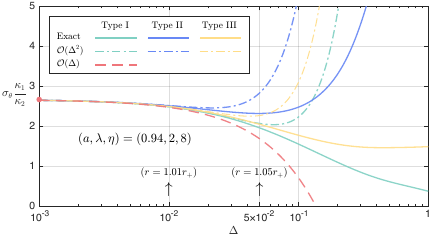}
	\includegraphics[width=3.2in]{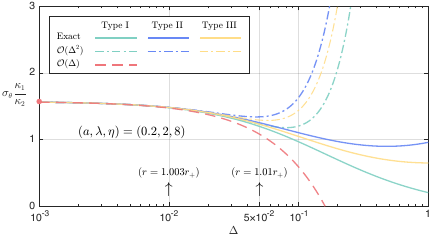}
	\centering
	\caption{Comparison of the ratio $z$ up to the next-to-leading order (Eq.~\eqref{master}) with the exact value under different inflow modes, under $a = 0.94$ (\textbf{Top}) and $a = 0.2$ (\textbf{Bottom}). We have chosen a typical direction for the emitted photons with $(\lambda, \eta) = (2, 8)$, and the field line angular velocity is set to $\Omega_B = 0.3 \Omega_H$. For results of other parameter choices, see Appendix.~\ref{appcompare}.}
	\label{test}
\end{figure}

In addition, the derivation of Eq.\eqref{efin} in the present study is based on a radial inflow, assuming $b^{\t} \propto u^{\t} = 0$, in accordance with MADs. 
However, we have numerically confirmed that introducing a small but nonzero vertical component preserves the leading-order result as given in Eq.\eqref{efin}, namely, 
 $z_0 = \sigma_\t \sqrt{\eta}\left(\lm - a\right)^{-1}$ still holds. However, the presence of $b^{\t}$ modifies the subleading-order term, and $u^{\t}$ enters into the expression for $z_1$, rendering it more complex. The full expression is omitted for brevity.

\section{NHP on the image plane}

\begin{figure}[ht!]
	\centering
	\includegraphics[width=3in]{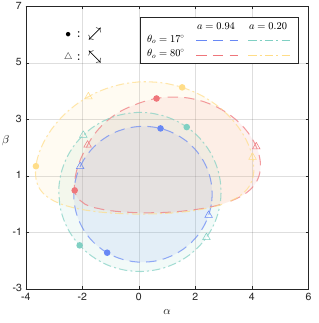}
	\centering
	\caption{The dashed lines represent the projected event horizons. The symbol ``$\bullet$" marks the locations where $\chi = -\pi/4$, and ``$\bigtriangleup$" marks the locations where $\chi = \pi/4$. The projection of the black hole's rotation direction onto the screen aligns with the $\beta$-axis.  
}
	\label{images}
\end{figure}

In Fig.~\ref{images}, we show the direct projection of the event horizon onto the image plane, together with the corresponding leading-order NHP pattern, i.e., $\chi = -\tan^{-1}\left(\b/\m\right) + \tan^{-1}z_0 $, for various black hole spins and inclination angles. For clarity, only those image points satisfying $\chi = \pm \pi/4$ are indicated, highlighting their dependence on both the spin and the inclination angle. Since $a$ and $\t_o$ affect the leading-order NHP in distinct and non-degenerate ways, their respective contributions can be disentangled from the EVPA distribution.

To better characterize the NHP pattern, we introduce the net EVPA: $\chi_{net} \equiv \chi - \varphi$, which indicates the angle between the polarization vector and the vector from the origin to the image point. Here, $\varphi =-  \tan^{-1}{(\a/\b)}$ is the polar angle on the image plane, measured with respect to the $\beta$-axis.
For small inclination angles, the polarization is nearly axisymmetric with respect to the line of sight, and we have $2\chi_{net} \approx \b_2$, where $\b_2$ is the second azimuthal Fourier mode of the linear polarization at fixed angular radius \citep{EventHorizonTelescope:2021srq,EventHorizonTelescope:2021bee, EHT:2024nwx}, serving as a good indicator of the overall polarization structure.

\begin{figure}[ht!]
	\centering
	\includegraphics[width=3.2in]{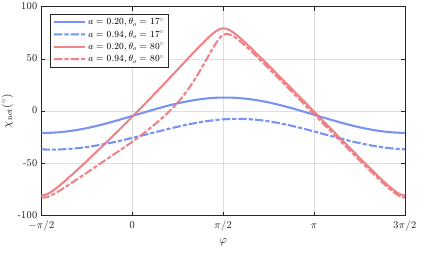}
	\centering
	\caption{The leading-order net EVPA along the projected event horizon as $\varphi$ changes, under different spins and inclination angles. We have taken the modulus of $\chi_{net}$ to bring it within the range of $[-\pi/2, \pi/2]$.}
	\label{chitheta}
\end{figure}

Fig.~\ref{chitheta} displays the variation of the net EVPA along the projected event horizon for different values of $a$ and $\t_o$. It is evident that the inclination angle has a significant impact on the variation of $\chi_{net}$ along the horizon contour, while increasing the spin results in an overall downward shift of $\chi_{net}$. For high inclination ($\t_o = 80^{\circ}$), $\chi_{net}$ in the lower half of the image plane is largely insensitive to the spin, whereas the upper half remains sensitive to changes in spin. In contrast, for low inclination ($\t_o = 17^{\circ}$), as favored by M87* shadow measurements \citep{EventHorizonTelescope:2019dse, EventHorizonTelescope:2019pgp, EventHorizonTelescope:2019ggy}, the entire NHP pattern is sensitive to the spin parameter: as $a$ increases from 0.2 to 0.94, $\chi_{net}$ shifts downward by nearly $20^{\circ}$. These results demonstrate that the NHP provides a potentially powerful and model-independent method for measuring black hole spin.

In the case of small inclination angles, Eq.\eqref{EVPAkerr} can be expanded in terms of $\t_o$ to yield an explicit expression for the NHP. In particular, for $\t_o = 0^{\circ}$, the leading-order net EVPA takes a particularly simple form: $\chi_{net} = -\tan^{-1} \left( a/\sqrt{(\rho_+)^2 - a^2}\right)$, where $\rho_+$ denotes the angular radius of the projected event horizon. 
Combining this formula with the semi-analytical relation $\rho_+ \approx 2\sqrt{r_+}$ \citep{Chael:2021rjo} yields the explicit expression for $\b_2$ given in Eq.~\eqref{NHPbeta}.
Further details are provided in  Appendix.~\ref{smallangle}.

\section{Summary and Discussion}

Assuming a stationary, axisymmetric, highly conducting accreting flow in the equatorial plane with polarized synchrotron emission, we explicitly showed that black hole spin imprints a characteristic polarization pattern via strong frame-dragging near the horizon. In the near-horizon expansion of the Kerr spacetime, the leading-order NHP depends only on the spacetime geometry and observer's inclination, remaining independent of the inflow structure or electron distribution. This property offers a novel and robust diagnostic of black hole spin. The next-to-leading-order term incorporates the field line angular velocity, indicating that polarimetric observations just outside the horizon can reveal black hole-magnetic field coupling and potentially the origin of jets. We further extended the analysis to the Kerr-Newman-Taub-NUT spacetime (Appendix~\ref{KNTNcase}), suggesting the broader applicability of the NHP formula.

Although derived under the assumptions of global stationarity and axisymmetry, the NHP formula Eq.\eqref{efin} characterizes the distribution of the near-horizon EVPA rather than total linear polarization degree. 
Thus, in turbulent accretion flows, the NHP can partially survive in the regions where turbulence is weak and the plasma and magnetic field remain locally coupled via Eq.\eqref{bmu}. 
Moreover, even if instantaneous turbulent flows may not display polarization patterns consistent with the NHP prediction, time-averaging can suppress fluctuations and recover underlying structures, offering a potential avenue for testing the formula in GRMHD simulations.
In the MAD regime, emissivity is greatly enhanced near the event horizon, counteracting gravitational redshift and suggesting that the NHP may be detectable with future high-resolution observations \citep{Ayzenberg:2023hfw, Johnson:2023ynn, Johnson:2024ttr, EventHorizonTelescope:2024whi}. 

The present study focused on accretion flow emission confined to the equatorial plane, excluding the contributions from off-equatorial regions such as the jet sheath. Our preliminary study suggests that for finite deviations from the equator, the NHP exhibits flow-mode dependence when the emitting region is geometrically thin. Extending the analysis to include jet emission is a goal for future work. Additionally, in the finite-conductivity regime, the coupling between flow velocity and magnetic field becomes more complex than Eq.~\eqref{bmu}, necessitating detailed plasma modeling, including the treatment of ion and electron drift currents--an issue we also intend to address in future studies.

We did not account for Faraday depolarization \citep{Ricarte:2020llx}. However, since the NHP formula is achromatic and purely geometric, and the rotation measure near M87* and Sgr A* is nearly constant, future broadband and multi-frequency polarimetric observations will allow effective separation of Faraday effects and recovery of the intrinsic NHP pattern. Additionally, at sufficiently high frequencies, Faraday rotation becomes negligible, further simplifying the interpretation.

\begin{acknowledgments}
The work is partly supported by NSFC Grant Nos. 12205013 and 12275004. YM is supported by the National Key R\&D Program of China (Grant No. 2023YFE0101200), the National Natural Science Foundation of China (Grant No. 12273022), and the Shanghai Municipality Orientation Program of Basic Research for International Scientists (Grant No. 22JC1410600).
\end{acknowledgments}

\appendix

\section{Magnetic field structure in the accreting plasma}\label{Appmag}

The electromagnetic field plays a crucial role in magnetohydrodynamics (MHD). The electric and magnetic fields measured in the plasma comoving frame are defined as
\bea
e^{\mu} = u_{\nu} F^{\mu\nu} \,, \quad b^{\mu} = -u_{\nu} (*F)^{\mu\nu} \,,
\eea
where where $F^{\m\n}$ is the electromagnetic tensor, $(*F)^{\mu\nu}$ is the dual tensor, and $u^{\m}$ is the flow four-velocity. Using the definitions of $e^\m$ and $b^\m$, the field strength tensor and its dual can be expressed as 
\bea \label{FEB}
F^{\mu\nu} = u^{\mu} e^{\nu} - u^{\nu} e^{\mu} + \epsilon^{\mu\nu\alpha\beta} u_{\alpha} b_{\beta} \,, \quad  
(*F)^{\mu\nu} = b^{\mu} u^{\nu} - b^{\nu} u^{\mu} - \epsilon^{\mu\nu\alpha\beta} e_{\alpha} u_{\beta} \,,
\eea
where $\epsilon^{\mu\nu\alpha\beta}$ is the Levi-Civita tensor in curved spacetime.
The electromagnetic energy-momentum tensor can be rewritten in terms of the electric and magnetic fields as \citep{McKinney:2006tf}:
\bea
T_{EM}^{\mu\nu} = \left( u^{\mu} u^{\nu} + \frac{1}{2} g^{\mu\nu} \right)(e^2 + b^2) - e^{\mu} e^{\nu} - b^{\mu} b^{\nu} \,.
\eea
In the study of GRMHD around black holes, it is convenient to introduce the so-called pseudo-electromagnetic fields, defined as the contractions of the electromagnetic tensor and dual tensor with the timelike Killing vector,
\bea\label{BLEM}
E^{\m} = (dt)_{\n} F^{\m\n} =  F^{t\m}  \,, \quad B^{\m} = -(dt)_{\n} (*F)^{\m\n} =  -(*F)^{t\m} \,.
\eea
These pseudo-fields do not belong to physical fields measured by any observer, but they are widely used in GRMHD due to their simplified form and computational convenience in stationary spacetimes. In fact, compared to the pseudo-fields, the electromagnetic fields measured in the spacetime normal observer differ only by a lapse function \citep{Komissarov:2004ms}. Using Eq.~\eqref{BLEM}, one can derive the transformation relations between the pseudo-fields and the magnetic field measured in the comoving frame:
\bea\label{transBb}
&&E^{\m} = F^{t\m} = - \epsilon^{t\m\a\b}b_{\a}u_{\b} \,, \quad B^{\m} = b^{[\m}u^{t]} \,, \\
&&b^{t} = g_{i\m} B^{i}u^{\m} \,, \quad  b^i = \f{B^i+ B^t u^i}{ u^t} \,.
\eea

In our study, we are interested in steady-state, axisymmetric plasma configurations embedded in a steady-state, axisymmetric spacetime. In spherical coordinates, the spacetime line element takes
\bea
\mathrm{d}s^2=g_{tt}dt^2+2g_{t\p}dtd\p + g_{\p\p}d\p^2 + g_{rr}dr^2 + g_{\t\t}d\t^2 \,. 
\eea
The physical quantities varies only in the $(r,\t)$ plane, and we henceforth denote this as the poloidal plane and refer to $r,\t$ as the poloidal coordinates. 
For accreting plasma with high enough conductivity, the electric field within the plasma frame is zero, $F^{\m\n}u_{\n} = 0$. This is also called the ideal MHD condition, widely used in both theoretical studies and numerical simulations. 
In the following, we would like to derive the field structure by imposing symmetry requirements and the ideal MHD condition. 
Since the electromagnetic tensor only depends on $r,\t$, the azimuthal component of electric field vanishes,
\bea
E^{\p} = F^{t\p}  = \left[ g^{tt}g^{\p\p}- (g^{t\p})^2 \right] F_{t\p} = 0 \,,
\eea 
where $F_{t\p}$ is evaluated as $F_{t\p} = \pl_t A_{\p} - \pl_{\p} A_t = 0$, with $A_{\m}(r,\t)$ the gauge potential. 
Applying the ideal MHD condition in the $t$ direction, we obtain the following orthogonality: 
\bea\label{EUortho}
 F^{ti}u_i = E^{P}u_{P} = 0\,.
\eea
Here and thereafter, we label the poloidal coordinates $r,\t$ by the index $P$. Meanwhile, the ideal MHD condition in the $\p$ (toroidal) direction leads to $F_{\p \m}u^{\m} = F_{\p P}u^{P} = 0$, which can be expressed in terms of $B^{P}$:
\bea\label{Bpoloidal}
 B^{r} u^{\t} -  B^{\t} u^{r} = 0 \,.
\eea
The above equation means that the magnetic field line is aligned with the streamline within the poloidal plane. The exceptional case is $u^r = u^{\t} = 0$, where the electromagnetic field cannot be determined by ideal MHD alone; additional conditions are required. For example, at the lauch point of a jet, the magnetic field is determined by the smoothness near the lauch point.   
Moreover, using the gauge potential, we have 
\bea\label{uA}
B^{r} = \f{ \pl_{\t}A_{\p} }{\sqrt{-g}} \,, \quad B^{\t} = -\f{\pl_{r}A_{\p}}{\sqrt{-g}} \,, \quad u^P \pl_P A_\p =0 \,,
\eea
where $g$ is the determinant of the spacetime metric.  The third equation comes from Eq.~\eqref{Bpoloidal},  meaning that $A_\p$ is conserved along the fluid streamlines. In fact, $A_{\p}(r,\t)$ marks the magnetic flux enclosed by a circular loop of radius $r$ and polar angle $\t$, and thus we denote it as the stream function of the magnetized plasma, $\psi \equiv A_{\p}(r,\t)$. 

The ideal MHD condition in the poloidal directions leads to
\bea\label{Epoloidal}
&&g_{rr} E^r = g^{tt} v^{\t} F_{r\t} + \left( g^{tt}v^{\p} - g^{t\p} \right) F_{r\p} \,, \nn \\
&&g_{\t\t} E^{\t} = g^{tt} v^{r} F_{\t r} + \left( g^{tt}v^{\p} - g^{t\p} \right) F_{\t\p} \,, 
\eea
where we have introduced $v^{\m} \equiv u^{\m}/u^t$. Multiply the first equation of Eq.~\eqref{Epoloidal} by $B^r$, and the second equation by $B^{\t}$, then add the two resulting equations together to obtain
\bea\label{EBortho}
g_{PP}E^PB^P = 0 \,,
\eea
hence the pseudo-electric field and pseudo-magnetic field are orthogonal within the poloidal plane. Moreover, applying the index lowering relation we have $g^{tt}F_{tP} = \left( g_{PP}E^P + g^{t\p}F_{P\p} \right)$. Combined with Eq.~\eqref{EBortho}, we can deduce  $F_{t\t}F_{r\p} = F_{tr}F_{\t\p}$. It is convenient to define the following ``angular velocity'':
\bea\label{defOmegaB}
\Omega_B \equiv \f{F_{tP}}{F_{P\p}} \,, \quad P = r, \t \,.
\eea
Note that this angular velocity does not characterize the local twisting of the field lines, i.e., the ratio of the toroidal to poloidal components of the magnetic field. Instead, it represents the global rotation of a magnetic field line around the black hole. Thus, $\Omega_B$ is often called the field line angular velocity. 
We can rewrite the pseudo-electric field as
\bea
E^{P} = - \left( \f{g^{PP}\sqrt{-g}}{g^{tt}} \right) \left(\Omega_B - \f{g^{t\p}}{g^{tt}} \right) \tilde{\epsilon}_{PP'} B^{P'} \,.
\eea
It becomes clear that the pseudo-electric field is induced by both the field line rotation and the spacetime frame dragging, represented by $g^{t\p}$. Then, we would like to explore the conservation law for $\Omega_B$. The bianchi ideneity of the electromagnetic field in the $\p$ direction gives $\pl_{r}F_{\t t} + \pl_{\t}F_{tr} = 0$ (In the stationary and axisymmetric case, $ \pl_{[r}F_{\t \p]} = \pl_{[t}F_{\t \p]} = \pl_{[t}F_{r \p]} = 0$ gives trivial result). Then using Eq.~\eqref{defOmegaB}, we arrive at 
\bea
F_{r\p} \pl_{\t}\Omega_B - F_{\t\p} \pl_{r}\Omega_B  + \left( \pl_{\t} F_{r\p} - \pl_r F_{\t\p}\right)\Omega_B  = 0 \,.
\eea 
Since $F_{P\p} = \pl_{P}\psi$, the last term on the left-hand side of the equation vanishes automatically. Therefore, we have 
\bea
B^P \pl_P \Omega_B = 0 \,,
\eea
which indicates that $\Omega_B$ is conserved along a field line within the poloidal plane. By using the definition of $\Omega_B$ and the ideal MHD condition in the poloidal directions, the toroidal component $B^{\p}$ can be written as
\bea\label{Bphi}
B^{\p} =  (u^{\p} - \Omega_B u^t)\f{B^P}{u^P}  \,. 
\eea

\begin{figure}[ht!]
	\centering
	\includegraphics[width=4.6in]{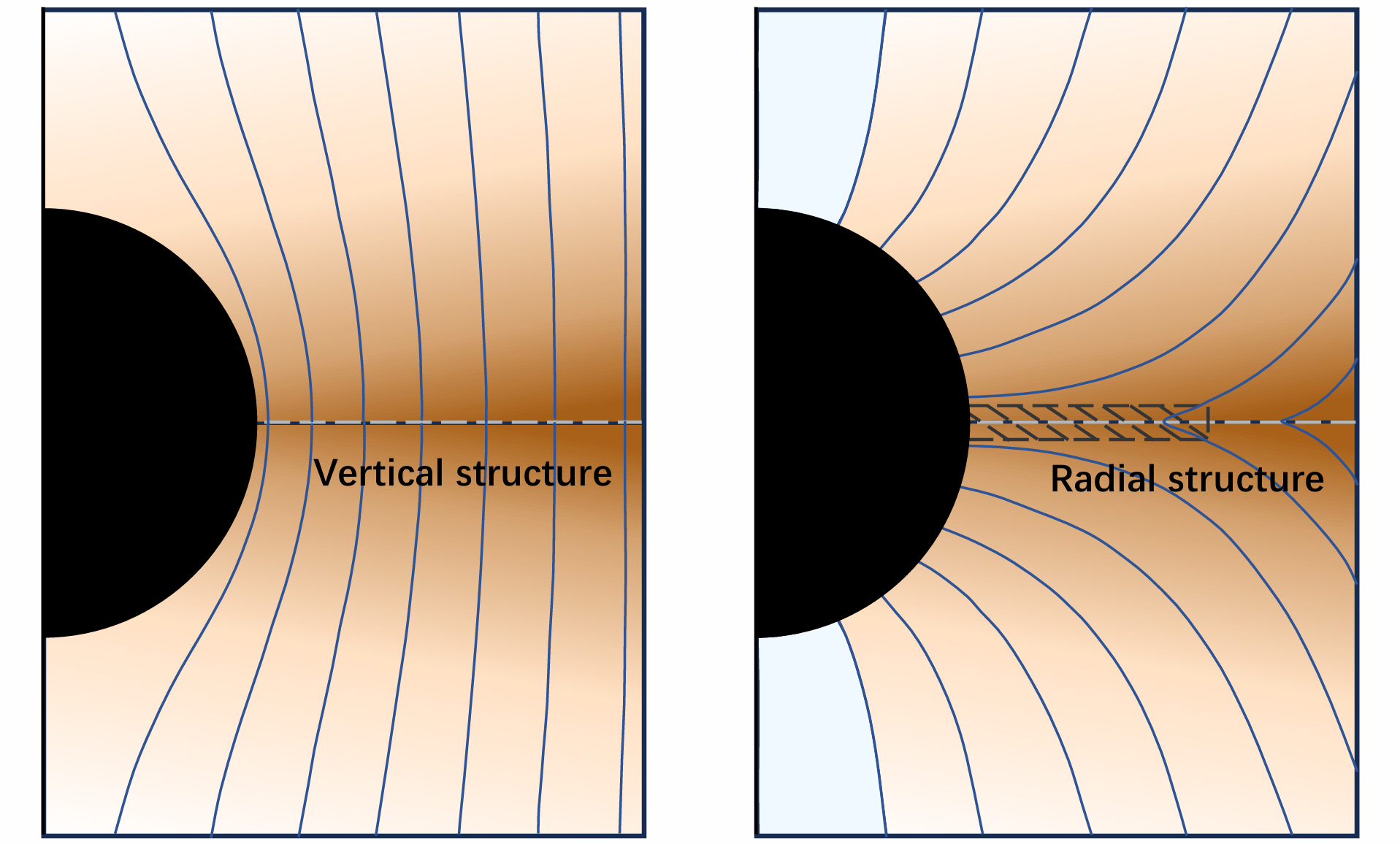}
	\centering
	\caption{Schematic diagram for the magnetic field topologies outside the black hole, within the poloidal plane. The dashed gray line denotes the location of the equatorial plane.  \textbf{Left:} vertical magnetic field across the equatorial plane. \textbf{Right:} Radial magnetic field. The shaded square region denotes the area of concentrated matter that predominantly contributes to the synchrotron radiation. The light blue regions near the polar axis indicate the jet regions of quasi-electromagnetic vacuum.}
	\label{illumag}
\end{figure}

Finally, we would like to consider the ideal-MHD-induced field close to the equatorial plane, aiming to capture the feature of the magnetic field within the accretion disk.  
Strictly at the equatorial plane, the condition $u^\t = 0$ holds to ensure a steady state without accumulation. If we further impose a reflection symmetry about the equatorial plane, the electromagnetic field at the equatorial plane satisfies $B^r = 0$ and $E^\t = 0$. Therefore, from Eq.~\eqref{Bpoloidal} we know that if $B^{\t} \neq 0$, then $u^r = 0$. In this case, from the expression of $B^{\p}$ (Eq.~\eqref{Bphi}) we find that
\bea
\Omega_B  = \f{u^{\p}}{u^t} = \Omega_{m} \,,
\eea
where $\Omega_{m}$ is the flow angular velocity. Thus, the vertical magnetic field lines are anchored in the equatorial matter and corotate with it. 
This condition is widely used in the study of jets induced by magnetized accretion flow, namely the Blanford-Payne jet \citep{Blandford:1982xxl}. The magnetic field measured in the plasma frame has a vertical structure in the poloidal plane,
\bea
b^{t} = B^{\p}_{\text{ver}} u_{\p} \,, \quad  b^{r} = 0 \,, \quad b^{\t} = \f{B^{\t}_{\text{ver}}}{u^t}\,, \quad b^{\p} = -B^{\p}_{\text{ver}}u_t \,,
\eea
where $B^{\t}_{\text{ver}}, B^{\p}_{\text{ver}}$ are undetermined functions of $r$. The left panel in Fig.~\ref{illumag} provides a representative structure of the vertical magnetic field. 
However, this vertical structure is not well applicable to the advection-dominated accretion flow very close to a supermassive back hole, where the accretion matter is dilute and inward flow is strong, with a large raidal velocity $u^r$.  

If $B^{\t} = 0$ at the equatorial plane, the accretion rate is no longer required to vanish. In this case, using Eq.~\eqref{EUortho} and Eq.~\eqref{Bphi}, we have $E^r = 0$ and $B^{\p} = 0$, respectively. Hence, for nonzero accretion rate, the electromagnetic field should be zero at the equatorial plane, $E^{\m}|_{\t= \pi/2} = B^{\m}|_{\t= \pi/2} = 0$. Slightly above/below the equatorial plane, we have $u^{\t},B^{\t},E^r \approx 0$, but $u^r, B^r, E^{\t} \neq 0$, corresponding to a split-monopole-like configuration near the equatorial plane, with a radial structure in the poloidal plane. With the expressions for the pseudo-magnetic field Eqs.~\eqref{uA}, \eqref{Bphi} and the relation Eq.~\eqref{transBb}, the magnetic field measured in the comoving frame takes the form of
\bea
b^{\m} = -\f{\pl_{\t}\psi}{\sqrt{-g}\,u^r} \left[ \left( u_t + \Omega_B u_{\p} \right)u^{\m} + \d^{\m}_t + \Omega_B\d^{\m}_{\p} \right] \, . 
\label{bmumu}
\eea
Very close to the equatorial plane, we have $u^{\m} \pl_{\m} \psi  \approx  u^r \pl_r \psi = 0$. Thus, $\psi$ is the function of $\t$ only. As long as the variation of $\psi$ with respect to $\t$ is not too sharp (a natural case, as is shown in \citep{Komissarov:2005wj}), we can safely replace $\pl_{\t}\psi$ by the averaged value,
\bea
\bar{\Psi} \equiv \f{1}{\d\t} \int^{\pi/2}_{\pi/2-\d \t} \pl_{\t}\psi d\t \, =\, \f{\psi(\pi/2+\d\t)- \psi(\pi/2)}{\d \t}  \,,
\eea
where $\d \t$ denotes the thickness of the accretion disk. 
For MAD simulation, the accretion matter is highly concentrated on the equatorial plane, see Figure.3 in \citep{Komissarov:2005wj}, Figure.2 in \citep{Chael:2018gzl}, Figure. 4 in \citep{Chael:2021rjo}, and we have $\d \t \ll \pi/2$. 
The solution Eq.~\eqref{bmumu} can well capture the magnetic field in the near-horizon part of MAD, as is demonstrated in various numerical simulations
\citep{Komissarov:2005wj,Tchekhovskoy:2011zx,mckinney2012general,Parfrey:2018dnc,Narayan:2021qfw}. 

In fact, for a given vertical magnetic field, the accretion process will generally leads to a split-monopole-like configuration near the black hole (more precisely, within the ergosphere).  
The right panel in Fig.~\ref{illumag} provides a representative magnetic field within simulated quasi-steady accretion flows. We see that far away from the black hole, the field line across the equatorial plane. Near the black hole, the field lines are connected to the event horizon and appear as a deformed split monopole. Close to the equatorial plane, we have $b^{\t}, B^{\t} \approx 0$, and the magnetic field is well described by Eq.~\eqref{bmumu}. At the eqautorial plane, a current sheet forms at the eqautorial plane, where magnetic reconnection occurs and the magnetic field structure gets turbulented \citep{Komissarov:2005wj}. Theoretically, this type of configuration genuinely  corresponds to the saturated state of magnetic flux threading the horizon. 

Magnetic field lines connected to the event horizon are dragged by the rotating black hole. In regions very close to the polar axis, the magnetic energy density is much higher than the plasma energy density, making the force-free approximation applicable \citep{thorne1982electrodynamics} (The light blue regions in Fig.~\ref{illumag}). In such cases, if the field line angular velocity $\Omega_B$ is less than the black hole's angular velocity $\Omega_H$, energy extraction can occur, resulting in the formation of an electromagnetic jet. In the jet region, $\Omega_B$ tends to approach $\Omega_H/2$ to enable efficient energy extraction \citep{Tchekhovskoy:2009ba}.

In regions near the equatorial plane where field lines still connect to the event horizon, $\Omega_B$ is jointly determined by the black hole’s spin and the inertia of the plasma, and typically lies between 0.2 and 0.5 \citep{Tchekhovskoy:2011zx}.
For field lines that do not connect to the event horizon but instead to the equatorial disk (such as all the field lines in Figure 1, or the outer ones in Figure 2), their angular velocity $\Omega_B$ is largely controlled by the plasma’s own rotation \citep{mckinney2012general}.

The rotation of magnetic field lines imprints characteristic signatures on the intrinsic polarization profiles of the emission, producing observable effects in accretion and jet polarimetry \citep{Gelles:2024tpz}. As demonstrated in the main text for the NHP pattern, the influence of $\Omega_B$ enters only at subleading order (see Eq.~\eqref{z0z1}), indicating that its variations are effectively suppressed at the horizon. Nevertheless, even slight departures from the horizon can induce polarization responses that are sensitive to changes in $\Omega_B$. Future high-precision observations may offer a means to distinguish different values of $\Omega_B$ within the emitting plasma.

\section{Examination of the applicability of the NHP formula}\label{appcompare}

The polarized emission profile, denoted by $f^{\m}_{e}$, is governed by several key physical parameters: the black hole spin $a$, the plasma inflow four-velocity $u^{\m}(r)$, magnetic field $b^{\m}(r)$, and the emission direction characterized by the impact parameters $\lm,\eta$. In a highly conducting plasma with a nonzero accretion rate, the magnetic field is not an independent degree of freedom but is instead constrained by the plasma motion and the field line angular velocity $\Omega_B$, as specified by Eq.~\eqref{bmumu}.
When this emission is projected onto the observer's image plane, it gives rise to a distribution of the EVPA. Utilizing the Penrose-Walker constant, the EVPA $\chi$ can be expressed as
\bea\label{generalchiKerr}
\chi\left(a,\t_o,\a,\b,\Omega_B,u^{\m}\right)  = \tan^{-1}\left(\f{\b}{\a+a \sin{\t_o}}\right) + \tan^{-1}z\left(a,\t_o,\a,\b,\Omega_B,u^{\m}\right) \,,
\eea
where $z$ denotes the ratio of the real to imaginary parts of the Penrose-Walker constant. Notably, plasmas characterized by different inflow profiles and field line angular velocities will exhibit distinct polarization signatures, primarily through the second term in Eq.~\eqref{generalchiKerr}. To compute the explicit value of the EVPA, one must first evaluate $z$ as a function of the impact parameters $\lm,\eta$, $r$ and the local plasma properties $u^{\m}(r),\Omega_B$ at the emission point. Then, we apply the lensing formalism in the Kerr spacetime, which provides an exact mapping between the emission position and the image-plane coordinates $(\a,\b)$ \citep{Gralla:2019drh}. Specifically, the impact parameters are directly related to $\a,\b$ by definition, while the radius of the emission point, encoded in the redshift factor $\D = (r-r_+)(r-r_-)$,  can be expressed as a function of $\lm,\eta$ using the so-called "inverse formula" \citep{Gralla:2019drh}.

Focusing on the near-horizon region, the main text demonstrates that the NHP pattern exhibits a remarkable insensitivity to variations in the inflow modes. In particular, the leading-order term in the near-horizon expansion of $z$ with respect to $\D$ is determined solely by the spacetime geometry, while the next-to-leading-order term depends only on the background spacetime and the angular velocity $\Omega_B$ (see Eqs.~\eqref{z0z1} and \eqref{efin}). 
To assess the convergence of the NHP approximation, we compare the flow-independent expression for the EVPA given in Eq.~\eqref{efin} with the EVPAs computed from several distinct inflow models. Specifically, we consider the following three representative inflow configurations:
(Type I) free-fall from rest at infinity, (Type II) plunging from the prograde ISCO, and (Type III) a sub-Keplerian inflow profile obtained from fitting GRMHD simulations in the MAD regime \citep{Chael:2021rjo}. The corresponding angular momentum density profiles are given by
\bea\label{threeinflow}
l(r) \equiv -\f{u_{\p}}{u_t} = \left\{
\begin{aligned}
	&\,\,0\,, \quad &\text{Type \uppercase\expandafter{\romannumeral1}} \\
	&\,\, \f{L_{ms} }{E_{ms} } \,, \quad &\text{Type \uppercase\expandafter{\romannumeral2}} \\
	&\,\,l_{ms} \sqrt{\f{r}{r_{ms}}} \,,  \quad &\text{Type \uppercase\expandafter{\romannumeral3}} 
\end{aligned} \right.
\eea
where $E_{ms}, L_{ms}$ and $r_{ms}$ denote the conserved energy, angular momentum, and radius, respectively, of particles orbiting at the prograde ISCO,
\bea
&&r_{ms} = 3+Z_2 -\sqrt{(3-Z_1)(3+Z_1 + 2Z_2)} \, , \nn \\
&&Z_1 = 1+ (1-a^2)^{1/3}\left[ (1+a)^{1/3} + (1-a)^{1/3} \right] \,, \quad Z_2 = \sqrt{3a^2+Z_1^2} \, , \nn \\
&&E_{ms} = \sqrt{ 1-\f{2}{3 r_{ms}} } \,, \, L_{ms} = 2\sqrt{3} \left( 1-\f{2a}{3\sqrt{r_{ms}}} \right) \,.
\eea
For accreting plasma of Type III, the radial inflow velocity is modeled by a broken power-law fitting function, motivated by simulation results \citep{Chael:2021rjo}:
\bea
v_r \equiv \f{u_r}{u_t} = v_{ms} \left( \f{r}{r_{ms}} \right)^{-6} \left[ \f{1}{2} + \f{1}{2} \left( \f{r}{r_{ms}} \right)^5 \right]^{4/5} \,.
\eea
To achieve the best agreement with the numerical results, the free parameters are set to $l_{ms} = 1$, $v_{ms} = 2$. It is worth noting that the inflow modes described above satisfy $Y = u_t+\Omega_H u_{\p} < 0$, which is a necessary criterion for accretion to occur.

The influence of the inflow mode is encapsulated in the second term of Eq.\eqref{generalchiKerr}. Therefore, to assess its impact, it suffices to compare the values of $z$ generated by the three inflow models in Eq.\eqref{threeinflow} with those predicted by the near-horizon expansion in Eq.~\eqref{master}.
This comparison can be carried out in two equivalent ways. The first approach is to directly compare the near-horizon expansion, $z_0(a,\t_o,\a,\b) + \D z_1(a,\t_o,\a,\b,\Omega_B)$, with the full values of $z(a,\t_o,\a,\b,\Omega_B,u^{\m})$ computed from different inflow models across image planes with varying inclination angles, black hole spins and field line angular velocities. 
In the polar coordinates $(\rho,\varphi)$, the convergence of Eq.~\eqref{master} can be quantified by $\d\rho \equiv \rho - \rho_+(\varphi)$, where $\rho_+(\varphi)$ denotes the angular radius of the projected event horizon.

Alternatively, since the mapping between $\{\lm,\eta,\D\}$ and $\{\a,\b\}$ is smooth and invertible outside the black hole for arbitrary values of $a$ and $\t_o$, the comparison can equivalently be performed at the emission point. By varying the emission direction $\lm,\eta$ freely—without imposing constraints on the number or location of observers—one can directly compare the near-horizon expansion of $z$ (Eq.~\eqref{master}) with the corresponding values derived from different inflow models. This method is more computationally convenient and enables a broader exploration of parameter space. Consequently, in the subsequent analysis, we focus on comparing $z$ at the emission point.
We have verified that the near-horizon expansion converges well with several representative emission directions, effectively capturing the overall convergence behavior of the NHP formula across the image plane. In what follows, we present results for two representative sets of impact parameters: $(\lm,\eta) = (2,8)$ and $(1,4)$.

\begin{figure}[ht!]
	\centering
	\includegraphics[width=7in]{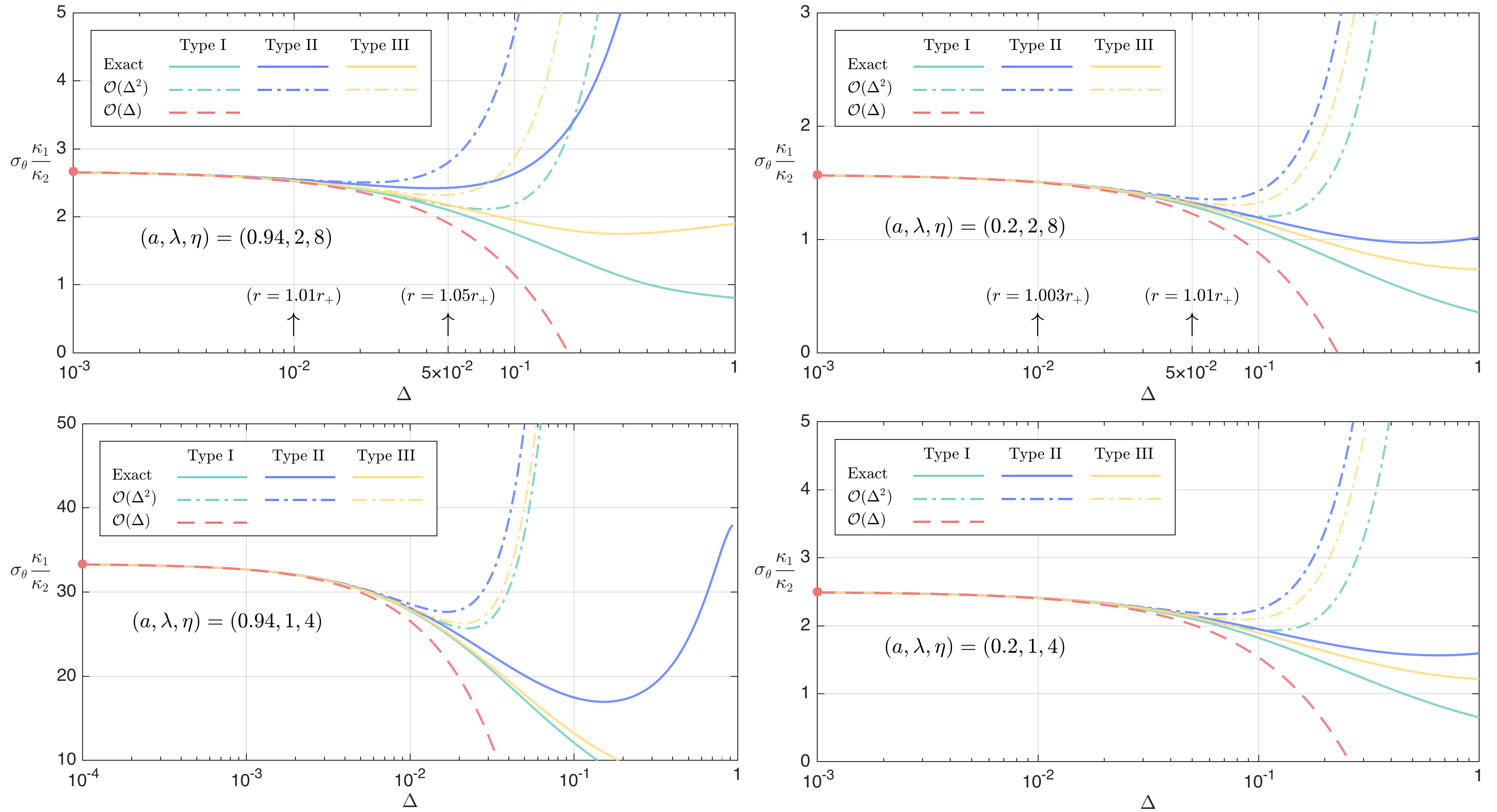}
	\centering
	\caption{Compare the universal relationship of $z$ with the exact $z$ for the plasma under three different flow conditions. The field line angular velocity is set to $\Omega_B = 0$.}
	\label{test02}
\end{figure}

\begin{figure}[ht!]
	\centering
		\includegraphics[width=7in]{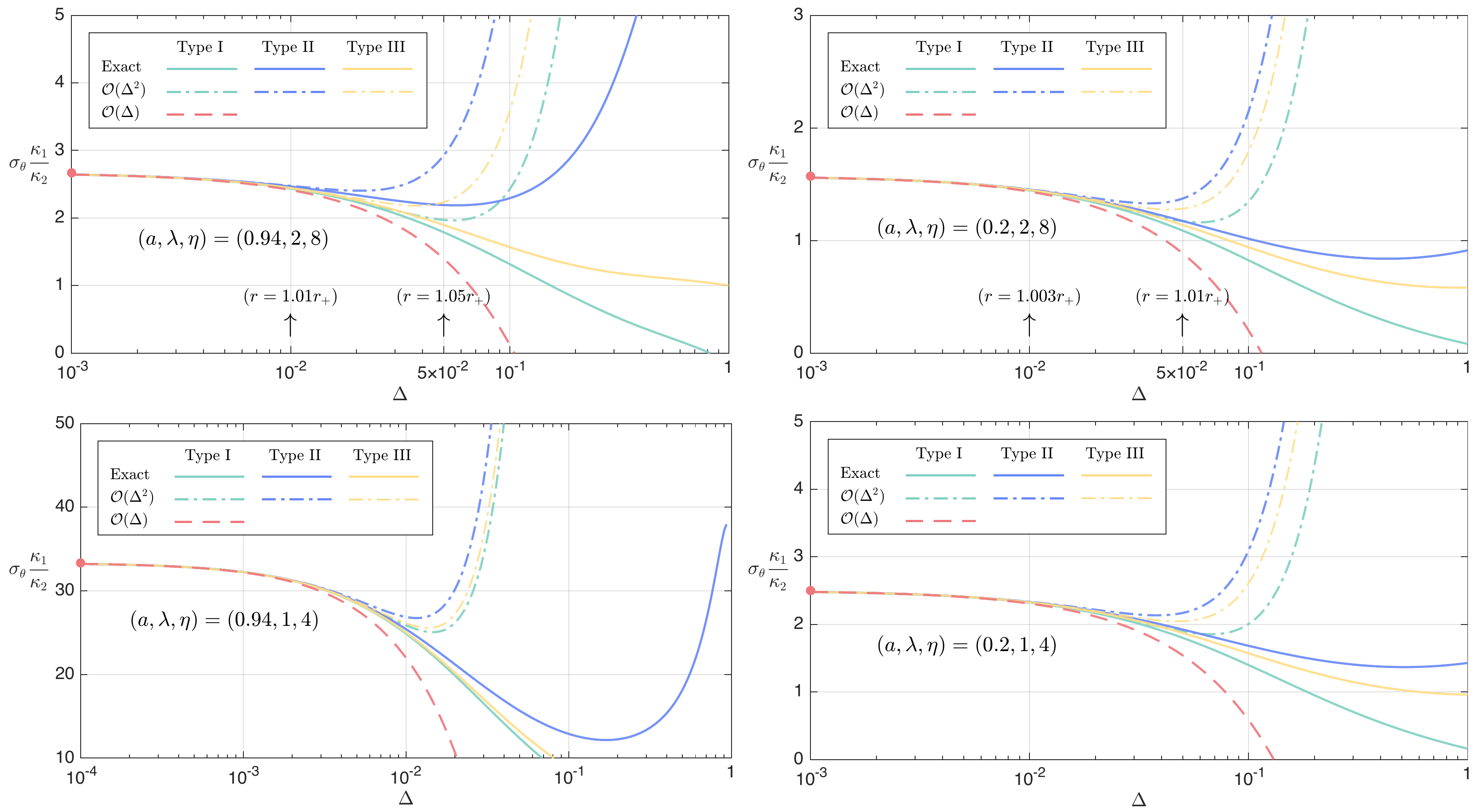}
	\centering
	\caption{Compare the universal relationship of $z$ with the exact $z$ for the plasma under three different flow conditions. The field line angular velocity is set to $\Omega_B = 0.5 \Omega_H$.}
	\label{test03}
\end{figure}

In Fig.\ref{test02}, we compare the near-horizon expansion given by Eq.\eqref{master} with the exact values of $z$ computed under the three inflow models, with the field line angular velocity fixed at $\Omega_B = 0$. 
The top row of Fig.\ref{test02} corresponds to photons emitted with impact parameters $(\lm,\eta) = (2,8)$. For a rapidly spinning black hole with $a = 0.94$, we find that the expansion up to next-to-leading order in Eq.\eqref{master} agrees remarkably well with the exact result within the range $ r = 1.01r_+ $. Beyond $r = 1.05r_+ $, the different inflow models begin to exhibit discernible deviations.
In contrast, for a slowly spinning black hole with $a = 0.2$, the same expansion already introduces noticeable discrepancies at $ r = 1.01r_+ $, indicating a slower convergence rate in the low-spin regime. The bottom row of Fig.~\ref{test02} shows results for photons with $(\lm,\eta) = (1,4)$.  In this case, the convergence radius for $a = 0.94$ is reduced compared to the $(\lm,\eta) = (2,8)$ case, while for $a = 0.2$, the convergence behavior remains largely unchanged.

Fig.~\ref{test03} presents the same comparison as Fig.\ref{test02}, except with the field line angular velocity set to $\Omega_B = 0.5 \,\Omega_H$. It is evident that the inclusion of a nonzero $\Omega_B$ does not qualitatively alter the behavior of $z$ as a function of $\D$ across different parameter choices. However, as demonstrated in Figs.~\ref{test},~\ref{test02}, and ~\ref{test03}, increasing $\Omega_B$ consistently leads to a reduction in the convergence radius of the near-horizon expansion. A quantitative understanding of this effect would require analyzing the influence of $\Omega_B$ on the next-to-next-to-leading order terms, which lies beyond the scope of the present work.

For completeness, we provide a few useful expressions related to the NHP pattern. To visualize the NHP structure on the image plane, it is convenient to switch to polar coordinates, $\a = -\rho \sin{\varphi}, \b = \rho\cos{\varphi}$. The corresponding relations for the impact parameters are $\lm = \rho\sin{\varphi}\sin{\t_o},\eta = (\cos^2{\varphi} + \sin^2{\varphi}\cos^2{\t_o})\rho^2 - a^2\cos^2{\t_o}$. The leading two terms in the near-horizon expansion of $z$ are given by
\bea
&&z_0 = \text{sign}(\cos{\t_o}) \f{ \sqrt{(\cos^2{\varphi} + \sin^2{\varphi}\cos^2{\t_o})\rho^2 -a^2\cos^2{\t_o}} }{a-\rho\sin{\varphi}\sin{\t_o}} \,, \nn \\
&&z_1 = \text{sign}(\cos{\t_o})\Omega_H^2 \, \f{(1-a\Omega_B)}{(\Omega_H -\Omega_B)} \, \f{ \left( \rho^2 + a^2\sin^2{\t_o} - 2a\rho\sin{\varphi}\sin{\t_o} \right) \sqrt{(\cos^2{\varphi} + \sin^2{\varphi}\cos^2{\t_o})\rho^2 - a^2\cos^2{\t_o}}}{2a^2 \left( a- \rho\sin{\varphi}\sin{\t_o} \right)^2 (1 - \Omega_H\rho\sin{\varphi}\sin{\t_o})}    \,,
\eea
Accordingly, when assessing the validity of the near-horizon expansion, one examines the behavior of $z$ as a function of the radial displacement $\d\rho = \rho - \rho_+(\varphi,a,\t_o)$, i.e., the distance from the projected event horizon $\rho = \rho_+(\varphi,a,\t_o)$ along a given direction $\varphi$. In \citep{Chael:2021rjo}, the authors presented a semi-analytic fit for the projected event horizon : 
\bea\label{horizonfit}
\rho_+(\varphi,a,\t_o) \approx 2\sqrt{r_+} - \left( 1 + \f{1}{2} \cos^2{\t_o} \right) \tan^{-1}\left( \cos{\varphi}\tan{\t_o} \right) \,.
\eea
The above formula is particularly accurate for observer's inclination angles less than $45^{\circ}$, where the projected event horizon deviates only mildly from a circle, and spin-induced asymmetries are not dominant. (It should be noted that in \citep{Chael:2021rjo}, $\varphi$ is defined as the angle relative to the $\a$-axis.)

\section{Explicit expressions  in the small inclination angle regime}\label{smallangle}

For an on-axis observer with inclination angle $\t_o = 0^{\circ}$, observables on the image plane depend solely on the angular radius $\rho$, and the polarization pattern exhibits axial symmetry about the line of sight (the spin axis). In this symmetric configuration, the net EVPA satisfies the relation 
$2\chi_{net}(\rho) = \b_2(\rho)$, where $\b_2(\rho)$ denotes the second azimuthal Fourier mode of the linear polarization at fixed angular radius \citep{EventHorizonTelescope:2021bee, EHT:2024nwx}. 

The observed polarization pattern arises from an interplay of several physical mechanisms, including the magnetic field structure, the plasma inflow mode, and the gravitational lensing. In the absence of plasma motion and gravitational effects on light propagation, a spinning black hole would produce a purely radial polarization pattern as seen by an on-axis observer. This stems from the fact that the emitted polarization vector is largely orthogonal to the local magnetic field, which, near the event horizon, adopts a perfectly toroidal configuration due to extreme frame dragging.

However, the relativistic motion of the accretion flow introduces significant modifications. The Lorentz boost from the plasma's comoving frame to the black hole frame induces aberration, causing the polarization vector of the emitted radiation to deviate from being strictly orthogonal to the magnetic field lines in the black hole frame. Simultaneously, frame dragging along photon trajectories leads to a non-trivial rotation of the polarization vector, a phenomenon commonly referred to as the gravitational Faraday effect \citep{Ishihara:1987dv}.

\begin{figure}[ht!]
	\centering
	\includegraphics[width=3.2in]{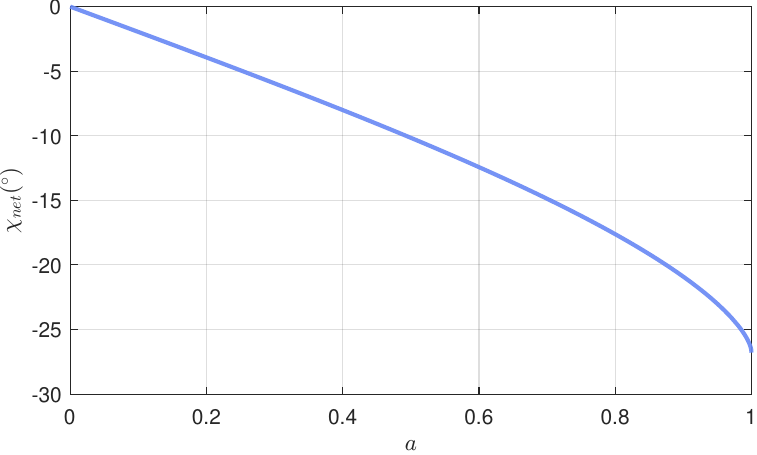}
	\centering
	\caption{The net EVPA at the event horizon in the images of a Kerr black hole with varying spins, viewed by an on-axis observer. We have taken the modulus of $\chi_{net}$ to bring it within the range of $[-\pi/2, \pi/2]$.}
	\label{chinet}
\end{figure}

In the near-horizon region, the aberration effect can be characterized by the spacetime geometry alone, as different inflow modes asymptotically converge to a universal toroidal structure, as shown in Eq.~\eqref{eou}. Together with the gravitational Faraday rotation, these two general relativistic effects imprint distinctive and robust signatures on the NHP pattern observed in the image plane.
The leading-order net EVPA takes a particularly simple form, as derived from Eqs.\eqref{z0z1} and \eqref{efin}:
\bea\label{faceonevpa}
\chi_{net} \, = \,-\tan^{-1}\left(\dfrac{a}{\sqrt{\rho_+(a)^2-a^2}} \right) 
\eea
where the angular radius of the projected event horizon is given approximately by the semi-analytic fit in Eq.\eqref{horizonfit}, $\rho_+(a) \approx 2\sqrt{r_+}$. 
Fig.~\ref{chinet} illustrates the dependence of the net EVPA, as given by Eq.~\eqref{faceonevpa}, on the black hole spin parameter, as seen by an on-axis observer. The net EVPA decreases monotonically with increasing spin, reflecting a transition in the polarization morphology—from a purely radial pattern in the non-rotating limit ($a \rightarrow 0$ but nonzero, $\chi_{net} \rightarrow 0$) to a clockwise spiral pattern in the extremal limit
($a \rightarrow 1$, $\chi_{net} \approx -26.8^{\circ}$).

For a small but nonzero inclination angle, we introduce the parameter $u \equiv \sin{\t_o} \ll 1$ to facilitate a perturbative expansion. In this limit, the impact parameters of light rays can be approximated as $\lm = \left(\rho \sin{\varphi}\right) u$, $\eta \approx \rho^2-a^2 + 2\left( a^2-\rho^2\sin^2{\varphi} \right) u $.
These expressions reveal a slight breaking of axial symmetry in the polarization pattern. For the leading-order EVPA, given by $\chi = -\tan^{-1}\left(\b/\m\right) + \tan^{-1}z_0\,$, we expand the net EVPA in powers of $u$, yielding
\bea\label{chiu}
\chi_{net}(\rho, \varphi)  = -\tan^{-1}{\dfrac{1}{z(\rho, \varphi)}}
 - \left(\f{a \cos{\varphi}}{\rho}\right) u  - \, \left(\f{a^2 \sin{2\varphi}}{2\rho^2}\right) u^2 + \mO(u^3) \, ,
\eea
where $z(\rho, \varphi) \approx z_0 + \D(\rho, \varphi) \, z_1$, and $z_0,z_1$ can be expanded as
\bea\label{zu}
z_0  &=& \, \f{\sqrt{\rho^2-a^2}}{a} + \left(\f{\rho\sqrt{\rho^2-a^2}\sin{\varphi}}{a^2}\right) u +\,\left[ \f{a^4 + \left( 2\rho^4-3a^2\rho^2 \right)\sin^2{\varphi}}{2a^3\sqrt{\rho^2-a^2}}\right] u^2+ \mO(u^3)\, . \\
f^{-1}_{\Omega} \, z_1 &=&  \f{\rho^2\sqrt{\rho^2-a^2}}{2a^4} - \sin{\varphi}\,\f{\rho\sqrt{\rho^2-a^2}}{4a^5} \left[ 4a^2 - \left( 5 - \sqrt{1-a^2}\rho^2 \right) \right]  u \nn \\
&+&  \left[ \left(18-a^2-6\sqrt{1-a^2}\right)\rho^4 + \left( a^4 -40a^2 + 10a^2 \sqrt{1-a^2}\right)\rho^2 + 4a^4 \left( 1-\sqrt{1-a^2} \right) \right] \f{\rho^2\sin^2{\varphi}}{8a^6\sqrt{\rho^2-a^2}}u^2 \nn \\
 &+& \left( 7\rho^2 - 4\rho^2 \cos{2\varphi} - 2a^2  \right)  \f{u^2}{4a^2\sqrt{\rho^2-a^2}} + \mO(u^3) \,,\label{z1u}
\eea
with $f_{\Omega} = \Omega_H^2 (1-a\Omega_B)(\Omega_H -\Omega_B)^{-1}$.
By retaining only the leading-order terms in Eq.\eqref{chiu}, we take the limit $z \rightarrow z_0$, where the asymmetry in the polarization pattern arises from the $\mO(u)$ contributions in Eqs.\eqref{chiu} and \eqref{zu}. 
For spin parameter $a = 0.94$ and inclination angle $\t_o = 17^{\circ} $, the maximal asymmetry occurs near $\varphi = -\pi/2, \pi/2$, where the difference in net EVPA reaches $\chi_{net}(\pi/2) - \chi_{net}(-\pi/2) \approx 30.72^{\circ}$, as illustrated in Fig.\ref{chitheta}. By imposing $\rho = \rho_{+}$ in Eq.\eqref{chiu} and substituting the semi-analytical fit for $\rho_+$, the behavior of the leading-order EVPA at small inclination angles can be fully determined. This explicit expression for the NHP pattern is particularly useful for constraining both the inclination angle and the black hole spin without degeneracy.

When the next-to-leading-order correction are included, i.e., $z \approx z_0 + \D z_1$, it becomes necessary to establish the explicit mapping between the emission radius $r = r_+ + \d r \approx r_+ + (r_+ - r_-)^{-1}\D  $ and the angular radius $\rho$. For an on-axis observer, this mapping can be approximated via the following expansion \citep{Gelles:2021kti}:
\bea\label{bigr}
&&r = \rho -1+\f{1-a^{2}}{2\rho}+\f{3(5\pi-16)}{4\rho^2}+\mathcal{O}(1/\rho^3) \,.
\eea
Although Eq.\eqref{bigr} is formally derived in the asymptotic limit $r_s \gg 1$, it remains a good approximation even in the near-horizon regime where $\d r \ll 1$ \citep{Gralla:2019drh}.
In particular, to $\mathcal{O}(\rho^{-1})$ accuracy, Eq.~\eqref{bigr} agrees remarkably well with the fitting formula $\rho_+ \approx 2\sqrt{r_+}$ proposed in \citep{Chael:2021rjo}, with the relative deviation below $0.2\%$. This implies that, in the face-on case with $\t_o = 0^{\circ}$, either Eq.\eqref{horizonfit} or Eq.~\eqref{bigr}, accurate to $\mathcal{O}(\rho^{-1})$, can be reliably used to evaluate $\rho_+$.

Therefore, by combining Eqs.\eqref{chiu}, \eqref{zu}, \eqref{z1u}, and \eqref{bigr}, one obtains an explicit expression for the near-horizon net EVPA accurate up to next-to-leading order. This result enables a direct and efficient analysis of the NHP pattern within a finite but small region outside the event horizon, rather than being restricted to the horizon itself.

Finally, it is important to emphasize that our analysis is applicable only to rotating black holes with nonzero spin. In the Schwarzschild case, the absence of frame dragging precludes both the development of an extreme toroidal plasma structure and the gravitational Faraday effect. Consequently, the polarization pattern in this case is governed by the inflow mode and the magnetic field configuration.

\section{NHP in the Kerr-Newman-Taub-NUT spacetime}\label{KNTNcase}

In future work, we aim to investigate whether the characteristic features of the NHP pattern persist in more general rotating black hole spacetimes. As a representative example, we consider a multi-parameter extension of the Kerr solution: the Kerr-Newman-Taub-NUT (KNTN) spacetime \citep{griffiths2009exact}. We apply the near-horizon analysis developed earlier to examine the structure of the NHP in this more general context.
Adopting the unit with $GM = c =1$, the KNTN spacetime is characterized by three parameters: the spin parameter $a$, the electric charge $Q$, and the NUT charge $l$, the latter of which introduces a nontrivial topological structure and breaks asymptotic flatness. In Boyer-Lindquist coordinates, the line element is given by
\begin{align}\label{metricKNTN}
\mathrm{d}s^2=&-\f{1}{\tilde{\Sigma}}\left(\tilde{\D}-a^2\sin^2{\t}\right)\mathrm{d}t^2+\frac{\tilde{\Sigma}}{\tilde{\D}}\mathrm{d}r^2+\tilde{\Sigma}\mathrm{d}\theta^2 \nn \\
&+\f{2}{\tilde{\Sigma}}\left[\tilde{\D} \mathcal{P}-a\left( \tilde{\Sigma}+a\mathcal{P} \right)\sin^2{\t} \right]  \mathrm{d}t\mathrm{d}\phi\nn \\ & + \f{1}{\tilde{\Sigma}} \left[ \left( \tilde{\Sigma}+a\mathcal{P} \right)^2\sin^2{\t}-\mathcal{P}^2\tilde{\D} \right] \mathrm{d}\phi^2  \,,
\end{align}
where $\tilde{\D} = (r^2-2r+a^2 +Q^2-l^2), \tilde{\Sigma} = r^2+\left( l + a\cos\t\right)^2 $, and $\mathcal{P} = a\sin^2{\t} - 2l \cos{\t}$. When $l = 0$, the spacetime reduces to the Kerr-Newman solution. For nonzero NUT charge, however, the spacetime is no longer asymptotically flat; in particular, the off-diagonal metric component $g_{t\p}$ asymptotically approaches $-2l\cos{\t}$, reflecting the presence of a gravitomagnetic monopole. To ensure the existence of an event horizon, the spacetime parameters must satisfy the condition $a^2+Q^2 \le 1+ l^2$. 
The event horizon is located at the outer root of the equation $\tilde{\D} = 0$ and is given by
\bea
\tilde{r}_+ = 1 + \sqrt{1-\left(a^2 + Q^2 - l^2\right)} \,.
\eea
To analyze light propagation in the KNTN spacetime, we consider the rescaled photon four-momentum associated with the metric in Eq.~\eqref{metricKNTN}. The energy-rescaled covariant four-momentum of a photon takes the form \citep{frolov2008higher, Grenzebach:2014fha}:
\bea\label{momentumKNTN}
&&p_{\m}\mathrm{d}x^{\m} = -\mathrm{d}t +\sigma_r \f{ \sqrt{\tilde{R}}}{\tilde{\D}}\mathrm{d}r +\sigma_{\t} \sqrt{\tilde{\T}}\mathrm{d}\t+ \lm \mathrm{d}\phi \, , \nn \\
&&\tilde{R}=(r^2+a^2+l^2-a\lm)^2-\tilde{\D} k \, ,  \nn \\
&&\tilde{\T} = k - \csc^2{\t}\left(\lm-a\sin^2{\t}+2l\cos{\t}\right)^2 \, ,
\eea
where $\lm, \eta$ are conserved impact parameters, and the constant $k$ is defined as $k = \eta + (\lm - a)^2$.  Furthermore, since the KNTN spacetime is of Petrov type D, it admits a nontrivial Killing spinor, which guarantees the existence of a conserved Penrose-Walker constant. This constant governs the parallel transport of the polarization vector along null geodesics. The Penrose-Walker constant takes the form
\bea \label{PWKNTN}
&& \,\kappa \, = \Xi \, (\mA-i\mB) \, , \nn \\
&&\mA= (p^{t} f^{r} - p^r f^t) + (a\sin^2{\t}-2l \cos{\t}) (p^r f^\p-p^\p f^r) \, , \nn \\
&&\mB = \left[(r^2+a^2+l^2) (p^{\p} f^{\t} - p^{\t} f^{\p}) -a (p^t f^\t - p^\t f^t)\right]\sin{\t} \, , \nn \\
&& \Xi = \left[r-i (l+a \cos{\t})\right]\left[-1+il+\f{Q^2}{r+i(l+a \cos{\t})}\right]^{-\f{1}{3}} \, .
\eea
Utilizing Eqs.\eqref{momentumKNTN} and \eqref{PWKNTN}, along with the general framework for an accreting magnetized plasma (Eqs.\eqref{bmu} and \eqref{fmu}), we compute the NHP in the KNTN spacetime in a manner analogous to that employed in the Kerr case. In the near-horizon limit $\D \rightarrow 0$, we expand the relevant physical quantities describing polarized radiation in powers of $\D$. Following a series of systematic calculations (details omitted here), we find that the near-horizon expansion of the ratio $\tilde{z} \equiv -\mA/\mB$ at the emission point exhibits precisely the same functional form as that in the Kerr spacetime,
\bea\label{masterKNTN}
\tilde{z} = \f{\sigma_{\t}\sqrt{\eta}}{\lm- a} \, + \left[\f{(1-a\Omega_B)}{(\Omega_h -\Omega_B)}\, \f{\sigma_{\t}\, \Omega_h^2\, k\,\sqrt{\eta}}{2a^2 (\lm-a)^2(\Omega_h \lm-1)} \right] \Delta  \, + \, \mO(\Delta^2) \,.
\eea
Here, the angular velocity of the black hole horizon is given by $\Omega_h = a (r_h^2+a^2+l^2)^{-1}$. Clearly, in the above expansion, the leading-order term depends solely on the spacetime geometry and the conserved impact parameters. The next-to-leading-order correction further incorporates the field line angular velocity $\Omega_B$, in addition to the spacetime geometry and impact parameters. Consequently, the resulting NHP pattern inherits the same qualitative features as in the Kerr spacetime: it is insensitive to the specific mode of plasma inflow and is governed only by the spacetime geometry and the magnetic field line rotation. 

\begin{figure}[ht!]
	\centering
	\includegraphics[width=7in]{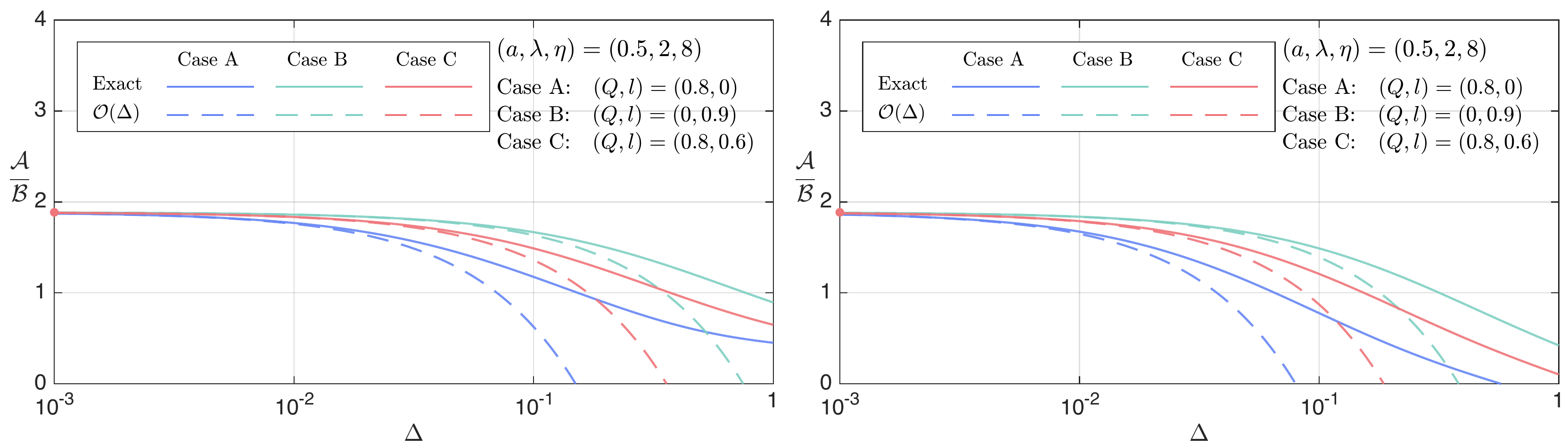}
	\centering
	\caption{Comparison of the ratio $\mA/\mB$ up to the next-to-leading order in Eq.~\eqref{masterKNTN} with the exact value for a free-falling plasma under different black hole parameters in the KNTN spacetime, with $\Omega_B = 0$ (\textbf{Left}) and $\Omega_B = 0.5\Omega_H$ (\textbf{Right}). We have chosen $(a, \lambda, \eta) = (0.5, 2, 8)$.}
	\label{KNTNtest}
\end{figure}

The expansion result in Eq.~\eqref{masterKNTN} is valid within the superradiant bound $\lm \leq 1/\Omega_h$, which ensures that the energy of photons emitted near the horizon remains positive. We also note the presence of poles in the expansion at $\lm = a$ and $\Omega_B = \Omega_h$, though a detailed analysis of these singularities is beyond the scope of this discussion.
To assess the validity and convergence domain of the near-horizon expansion, we compare the analytic result from Eq.\eqref{masterKNTN} with the exact numerical values of $\tilde{z}$ under a specific inflow mode, as shown in Fig.\ref{KNTNtest}. We see that the expansion, truncated at next-to-leading order in $\D$, remains accurate within the regime $0 \leq \D \lesssim 10^{-1}$.

Next, we propagate the near-horizon emission to the observer’s sky in order to explicitly determine the NHP pattern. Although the presence of the NUT charge renders the spacetime non-asymptotically flat, it is still possible to define a locally inertial observer by constructing an orthonormal tetrad at the distance. Following the standard procedure for setting up a locally nonrotating frame and defining photon arrival coordinates \citep{cunningham1973optical}, we introduce the coordinates $(\a, \b)$ to parametrize the image plane:
\bea
\a = -\f{\lm}{\sin{\t_o}} \, , \quad \b = \pm_{\b} \sqrt{k - \csc^2{\t_o}\left(\lm-a\sin^2{\t_o}+2l\cos{\t_o}\right)^2} \,,
\eea
with $\t_o$ being the observer's inclination angle. At large distances from the black hole, the radial component of the incoming photon's four-momentum dominates, satisfying $p^r \gg rp^{\t},rp^{\p}$, and we can normalize the photon's four-momentum such that $-p_t = p^t = 1, p_r = p^r = 1$. The components of the polarization vector projected onto the image plane axes are then given by
\bea\label{KNTNfo}
f^{\a}_{(o)} = \f{(f_{(o)})_{\p}}{r_o\sin{\t_o}} = r_o\sin{\t_o}f^{\p}_{(o)} \, , \quad f^{\b}_{(o)} = -\f{(f_{(o)})_{\t}}{r_o} = -r_o f^{\t}_{(o)} \, ,
\eea
where $r_o \gg 1$ denotes the radial coordinate of the distant observer. Imposing the orthogonality condition $p_{\m}f^{\m} = 0$ then yields $f^r = f^t - \b f^{\t} - \lm f^{\p}$. 
Hence, at the location of the distant observer, the Penrose-Walker constant simplifies to $\kappa = r_o \left(il-1 \right)^{-1/3}\left( \mA_o + i \mB_o\right)$, 
\bea
&&\mA_o =  -\b f^{\t} - \m\sin{\t_o} f^{\phi} = -\f{1}{r_o} \left( \m f^{\a}_{(o)} - \b f^{\b}_{(o)} \right) \,, \nn \\
&&\mB_o =  \m f^{\t} - \b \sin{\t_o} f^{\phi} = -\f{1}{r_o} \left( \b f^{\a}_{(o)} + \m f^{\b}_{(o)} \right) \,,
\eea
where $\m = -(\a + a \sin{\t_o}-2l\cot{\t_o})$. The EVPA of arrivaling photons is obtained through
\bea\label{evpa1}
\chi \equiv  -\tan^{-1}{\f{f^{\a}_{(o)}}{f^{\b}_{(o)}}} = \tan^{-1}\left(\frac{\mu \tilde{z}_o- \beta }{\beta \tilde{z}_o + \mu}\right)
\,, \quad \tilde{z}_o \equiv -\f{\mA_o}{\mB_o} \,.
\eea
Through Eq.~\eqref{PWKNTN}, $(\mA_o, \mB_o)$ can be related to the near-horizon components $(\mA, \mB)$ via a reflection matrix that depends solely on the emission position and the observer's radius. Assuming the photon is emitted from the equatorial plane, this relation takes the form:
\bea\label{matrix}
&& \zeta _o 
\left(\begin{matrix}
\cos{\d_o} & \sin{\d_o} \\
\sin{\d_o} &-\cos{\d_o}
\end{matrix}\right) \left(\begin{matrix}
\mA_o  \\
\mB_o
\end{matrix}\right)  = \, 
 \zeta
 \left(\begin{matrix}
\cos{\delta} & \sin{\delta} \\
\sin{\delta} &-\cos{\delta}
\end{matrix}\right) \left(\begin{matrix}
\mA  \\
\mB
\end{matrix}\right) \, ,
\eea
where 
\bea
&& \zeta _o = r_o \left(l^2+1\right)^{-\f{1}{6}} \,, \quad \, \d_o = \f{\tan^{-1}{l}}{3}\,, \nn \\
&& \zeta = \sqrt{r^2+l^2}\left[ \left( 1-\f{rQ^2}{r^2+l^2} \right)^2 +\left( l-\f{lQ^2}{r^2+l^2} \right)^2  \right]^{-\f{1}{6}} \, , \nn \\
&&\delta = -\tan^{-1} {\left(\f{l}{r}\right)}  + \f{1}{3} \tan^{-1}{\left[ \f{l(r^2+l^2-Q^2)}{r^2-Q^2r+l^2} \right]} \, .
\eea
Therefore, combining Eqs.~\eqref{masterKNTN}, \eqref{evpa1}, and \eqref{matrix}, we find that, to leading order in $\D$, the near-horizon polarization angle $\chi$ is independent of the plasma properties and instead encodes purely the spacetime geometry. This suggests that the polarization pattern near the horizon reflects intrinsic geometrical features of the spacetime and may be a universal characteristic of Petrov Type D spacetimes. The polarization angle is given by
\bea\label{NHPKNTN}
 \chi =-\tan^{-1}\left(\f{\b}{\m}\right) + \tan^{-1}\tilde{z}_o  \,, \quad \tilde{z}_o =   \f{\tilde{z}\cos{(\d + \d_o)}  - \sin{(\d + \d_o)}}{\tilde{z}\sin{(\d - \d_o)} - \cos{(\d - \d_o)}} \,.
\eea
Finally, by substituting Eq.\eqref{masterKNTN} into Eq.\eqref{NHPKNTN} and expanding $\d$ in powers of $\D$ in Eq.~\eqref{NHPKNTN}, one can obtain the explicit near-horizon expansion of the EVPA in the KNTN spacetime. As the derivation is not particularly illuminating, we omit the detailed steps here.

\bibliography{refs}

\begin{thebibliography}{}
\expandafter\ifx\csname natexlab\endcsname\relax\def\natexlab#1{#1}\fi
\providecommand{\url}[1]{\href{#1}{#1}}
\providecommand{\dodoi}[1]{doi:~\href{http://doi.org/#1}{\nolinkurl{#1}}}
\providecommand{\doeprint}[1]{\href{http://ascl.net/#1}{\nolinkurl{http://ascl.net/#1}}}
\providecommand{\doarXiv}[1]{\href{https://arxiv.org/abs/#1}{\nolinkurl{https://arxiv.org/abs/#1}}}

\bibitem[{Ayzenberg {et~al.}(2023)}]{Ayzenberg:2023hfw}
Ayzenberg, D., {et~al.} 2023.
\newblock \doarXiv{2312.02130}

\bibitem[{Bardeen(1973)}]{Bardeen:1973tla}
Bardeen, J.~M. 1973, Proceedings, Ecole d'Et\'e de Physique Th\'eorique: Les
  Astres Occlus : Les Houches, France, August, 1972, 215-240, 215

\bibitem[{Beckwith \& Done(2005)}]{Beckwith:2004ae}
Beckwith, K., \& Done, C. 2005, Mon. Not. Roy. Astron. Soc., 359, 1217,
  \dodoi{10.1111/j.1365-2966.2005.08980.x}

\bibitem[{Blandford \& Payne(1982)}]{Blandford:1982xxl}
Blandford, R.~D., \& Payne, D.~G. 1982, Mon. Not. Roy. Astron. Soc., 199, 883,
  \dodoi{10.1093/mnras/199.4.883}

\bibitem[{Blandford \& Znajek(1977)}]{Blandford:1977ds}
Blandford, R.~D., \& Znajek, R.~L. 1977, Mon. Not. Roy. Astron. Soc., 179, 433,
  \dodoi{10.1093/mnras/179.3.433}

\bibitem[{Carter(1968)}]{Carter:1968rr}
Carter, B. 1968, Phys. Rev., 174, 1559, \dodoi{10.1103/PhysRev.174.1559}

\bibitem[{Chael {et~al.}(2021)Chael, Johnson, \& Lupsasca}]{Chael:2021rjo}
Chael, A., Johnson, M.~D., \& Lupsasca, A. 2021, Astrophys. J., 918, 6,
  \dodoi{10.3847/1538-4357/ac09ee}

\bibitem[{Chael {et~al.}(2023)Chael, Lupsasca, Wong, \&
  Quataert}]{Chael:2023pwp}
Chael, A., Lupsasca, A., Wong, G.~N., \& Quataert, E. 2023, Astrophys. J., 958,
  65, \dodoi{10.3847/1538-4357/acf92d}

\bibitem[{Chael {et~al.}(2019)Chael, Narayan, \& Johnson}]{Chael:2018gzl}
Chael, A., Narayan, R., \& Johnson, M.~D. 2019, Mon. Not. Roy. Astron. Soc.,
  486, 2873, \dodoi{10.1093/mnras/stz988}

\bibitem[{Chandrasekhar(1985)}]{Chandrasekhar:1985kt}
Chandrasekhar, S. 1985, {The mathematical theory of black holes}

\bibitem[{Comp\`ere {et~al.}(2022)Comp\`ere, Liu, \& Long}]{Compere:2021bkk}
Comp\`ere, G., Liu, Y., \& Long, J. 2022, Phys. Rev. D, 105, 024075,
  \dodoi{10.1103/PhysRevD.105.024075}

\bibitem[{Cunningham \& Bardeen(1973)}]{cunningham1973optical}
Cunningham, C.~T., \& Bardeen, J.~M. 1973, Astrophysical Journal, Vol. 183, pp.
  237-264 (1973), 183, 237

\bibitem[{Dokuchaev \& Nazarova(2020)}]{Dokuchaev:2019jqq}
Dokuchaev, V.~I., \& Nazarova, N.~O. 2020, Usp. Fiz. Nauk, 190, 627,
  \dodoi{10.3367/UFNe.2020.01.038717}

\bibitem[{Emami {et~al.}(2023)}]{Emami:2022kci}
Emami, R., {et~al.} 2023, Astrophys. J., 950, 38,
  \dodoi{10.3847/1538-4357/acc8cd}

\bibitem[{{{Event Horizon Telescope
  Collaboration}}(2019{\natexlab{a}})}]{EventHorizonTelescope:2019dse}
{{Event Horizon Telescope Collaboration}}. 2019{\natexlab{a}}, Astrophys. J.
  Lett., 875, L1, \dodoi{10.3847/2041-8213/ab0ec7}

\bibitem[{{{Event Horizon Telescope
  Collaboration}}(2019{\natexlab{b}})}]{EventHorizonTelescope:2019pgp}
---. 2019{\natexlab{b}}, Astrophys. J. Lett., 875, L5,
  \dodoi{10.3847/2041-8213/ab0f43}

\bibitem[{{{Event Horizon Telescope
  Collaboration}}(2019{\natexlab{c}})}]{EventHorizonTelescope:2019ggy}
---. 2019{\natexlab{c}}, Astrophys. J. Lett., 875, L6,
  \dodoi{10.3847/2041-8213/ab1141}

\bibitem[{{{Event Horizon Telescope
  Collaboration}}(2021{\natexlab{a}})}]{EventHorizonTelescope:2021bee}
---. 2021{\natexlab{a}}, Astrophys. J. Lett., 910, L12,
  \dodoi{10.3847/2041-8213/abe71d}

\bibitem[{{{Event Horizon Telescope
  Collaboration}}(2021{\natexlab{b}})}]{EventHorizonTelescope:2021srq}
---. 2021{\natexlab{b}}, Astrophys. J. Lett., 910, L13,
  \dodoi{10.3847/2041-8213/abe4de}

\bibitem[{{{Event Horizon Telescope
  Collaboration}}(2022{\natexlab{a}})}]{EventHorizonTelescope:2022wkp}
---. 2022{\natexlab{a}}, Astrophys. J. Lett., 930, L12,
  \dodoi{10.3847/2041-8213/ac6674}

\bibitem[{{{Event Horizon Telescope
  Collaboration}}(2022{\natexlab{b}})}]{EventHorizonTelescope:2022urf}
---. 2022{\natexlab{b}}, Astrophys. J. Lett., 930, L16,
  \dodoi{10.3847/2041-8213/ac6672}

\bibitem[{{{Event Horizon Telescope
  Collaboration}}(2024{\natexlab{a}})}]{EHT:2024nwx}
---. 2024{\natexlab{a}}, Astrophys. J. Lett., 964, L25,
  \dodoi{10.3847/2041-8213/ad2df0}

\bibitem[{{{Event Horizon Telescope
  Collaboration}}(2024{\natexlab{b}})}]{EHT:2024ehx}
---. 2024{\natexlab{b}}, Astrophys. J. Lett., 964, L26,
  \dodoi{10.3847/2041-8213/ad2df1}

\bibitem[{{{Event Horizon Telescope
  Collaboration}}(2024{\natexlab{c}})}]{EventHorizonTelescope:2024whi}
---. 2024{\natexlab{c}}.
\newblock \doarXiv{2410.02986}

\bibitem[{Frolov \& Kubiz{\v{n}}{\'a}k(2008)}]{frolov2008higher}
Frolov, V.~P., \& Kubiz{\v{n}}{\'a}k, D. 2008, Classical and Quantum Gravity,
  25, 154005

\bibitem[{Gelles {et~al.}(2025)Gelles, Chael, \& Quataert}]{Gelles:2024tpz}
Gelles, Z., Chael, A., \& Quataert, E. 2025, Astrophys. J., 981, 204,
  \dodoi{10.3847/1538-4357/adb1aa}

\bibitem[{Gelles {et~al.}(2021)Gelles, Himwich, Palumbo, \&
  Johnson}]{Gelles:2021kti}
Gelles, Z., Himwich, E., Palumbo, D. C.~M., \& Johnson, M.~D. 2021, Phys. Rev.
  D, 104, 044060, \dodoi{10.1103/PhysRevD.104.044060}

\bibitem[{Gralla \& Lupsasca(2020)}]{Gralla:2019drh}
Gralla, S.~E., \& Lupsasca, A. 2020, Phys. Rev. D, 101, 044031,
  \dodoi{10.1103/PhysRevD.101.044031}

\bibitem[{Gralla {et~al.}(2020)Gralla, Lupsasca, \& Marrone}]{Gralla:2020srx}
Gralla, S.~E., Lupsasca, A., \& Marrone, D.~P. 2020, Phys. Rev. D, 102, 124004,
  \dodoi{10.1103/PhysRevD.102.124004}

\bibitem[{Grenzebach {et~al.}(2014)Grenzebach, Perlick, \&
  L\"ammerzahl}]{Grenzebach:2014fha}
Grenzebach, A., Perlick, V., \& L\"ammerzahl, C. 2014, Phys. Rev. D, 89,
  124004, \dodoi{10.1103/PhysRevD.89.124004}

\bibitem[{Griffiths \& Podolský(2009)}]{griffiths2009exact}
Griffiths, J.~B., \& Podolský, J. 2009, Exact space-times in Einstein's
  general relativity (Cambridge University Press)

\bibitem[{Hadar {et~al.}(2021)Hadar, Johnson, Lupsasca, \&
  Wong}]{Hadar:2020fda}
Hadar, S., Johnson, M.~D., Lupsasca, A., \& Wong, G.~N. 2021, Phys. Rev. D,
  103, 104038, \dodoi{10.1103/PhysRevD.103.104038}

\bibitem[{Himwich {et~al.}(2020)Himwich, Johnson, Lupsasca, \&
  Strominger}]{Himwich:2020msm}
Himwich, E., Johnson, M.~D., Lupsasca, A., \& Strominger, A. 2020, Phys. Rev.
  D, 101, 084020, \dodoi{10.1103/PhysRevD.101.084020}

\bibitem[{Hou {et~al.}(2023)Hou, Zhang, Guo, \& Chen}]{Hou:2023bep}
Hou, Y., Zhang, Z., Guo, M., \& Chen, B. 2023.
\newblock \doarXiv{2309.13304}

\bibitem[{Ishihara {et~al.}(1988)Ishihara, Takahashi, \&
  Tomimatsu}]{Ishihara:1987dv}
Ishihara, H., Takahashi, M., \& Tomimatsu, A. 1988, Phys. Rev. D, 38, 472,
  \dodoi{10.1103/PhysRevD.38.472}

\bibitem[{Jia {et~al.}(2024)Jia, Quataert, Lupsasca, \& Wong}]{Jia:2024mlb}
Jia, H., Quataert, E., Lupsasca, A., \& Wong, G.~N. 2024.
\newblock \doarXiv{2405.08804}

\bibitem[{Johnson {et~al.}(2023)}]{Johnson:2023ynn}
Johnson, M.~D., {et~al.} 2023, Galaxies, 11, 61,
  \dodoi{10.3390/galaxies11030061}

\bibitem[{Johnson {et~al.}(2024)}]{Johnson:2024ttr}
---. 2024.
\newblock \doarXiv{2406.12917}

\bibitem[{Komissarov(2004)}]{Komissarov:2004ms}
Komissarov, S.~S. 2004, Mon. Not. Roy. Astron. Soc., 350, 407,
  \dodoi{10.1111/j.1365-2966.2004.07446.x}

\bibitem[{Komissarov(2005)}]{Komissarov:2005wj}
---. 2005, Mon. Not. Roy. Astron. Soc., 359, 801,
  \dodoi{10.1111/j.1365-2966.2005.08974.x}

\bibitem[{Lockhart \& Gralla(2022)}]{Lockhart:2022rui}
Lockhart, W., \& Gralla, S.~E. 2022, Mon. Not. Roy. Astron. Soc., 517, 2462,
  \dodoi{10.1093/mnras/stac2743}

\bibitem[{McKinney(2006)}]{McKinney:2006tf}
McKinney, J.~C. 2006, Mon. Not. Roy. Astron. Soc., 368, 1561,
  \dodoi{10.1111/j.1365-2966.2006.10256.x}

\bibitem[{McKinney {et~al.}(2012)McKinney, Tchekhovskoy, \&
  Blandford}]{mckinney2012general}
McKinney, J.~C., Tchekhovskoy, A., \& Blandford, R.~D. 2012, Monthly Notices of
  the Royal Astronomical Society, 423, 3083

\bibitem[{Medeiros {et~al.}(2022)Medeiros, Chan, Narayan, Ozel, \&
  Psaltis}]{Medeiros:2021apx}
Medeiros, L., Chan, C.-K., Narayan, R., Ozel, F., \& Psaltis, D. 2022,
  Astrophys. J., 924, 46, \dodoi{10.3847/1538-4357/ac33a7}

\bibitem[{Moscibrodzka {et~al.}(2017)Moscibrodzka, Dexter, Davelaar, \&
  Falcke}]{Moscibrodzka:2017gdx}
Moscibrodzka, M., Dexter, J., Davelaar, J., \& Falcke, H. 2017, Mon. Not. Roy.
  Astron. Soc., 468, 2214, \dodoi{10.1093/mnras/stx587}

\bibitem[{Narayan {et~al.}(2022)Narayan, Chael, Chatterjee, Ricarte, \&
  Curd}]{Narayan:2021qfw}
Narayan, R., Chael, A., Chatterjee, K., Ricarte, A., \& Curd, B. 2022, Mon.
  Not. Roy. Astron. Soc., 511, 3795, \dodoi{10.1093/mnras/stac285}

\bibitem[{Narayan {et~al.}(2003)Narayan, Igumenshchev, \&
  Abramowicz}]{Narayan:2003by}
Narayan, R., Igumenshchev, I.~V., \& Abramowicz, M.~A. 2003, Publ. Astron. Soc.
  Jap., 55, L69, \dodoi{10.1093/pasj/55.6.L69}

\bibitem[{Palumbo(2025)}]{Palumbo:2024czv}
Palumbo, D. C.~M. 2025, Astrophys. J. Lett., 978, L4,
  \dodoi{10.3847/2041-8213/ad9bb4}

\bibitem[{Palumbo {et~al.}(2024)Palumbo, Baubock, \& Gammie}]{Palumbo:2024jtz}
Palumbo, D. C.~M., Baubock, M., \& Gammie, C.~F. 2024, Astrophys. J., 970, 151,
  \dodoi{10.3847/1538-4357/ad5fed}

\bibitem[{Palumbo {et~al.}(2020)Palumbo, Wong, \& Prather}]{Palumbo:2020flt}
Palumbo, D. C.~M., Wong, G.~N., \& Prather, B.~S. 2020, Astrophys. J., 894,
  156, \dodoi{10.3847/1538-4357/ab86ac}

\bibitem[{Parfrey {et~al.}(2019)Parfrey, Philippov, \&
  Cerutti}]{Parfrey:2018dnc}
Parfrey, K., Philippov, A., \& Cerutti, B. 2019, Phys. Rev. Lett., 122, 035101,
  \dodoi{10.1103/PhysRevLett.122.035101}

\bibitem[{Petrov(1954)}]{petrov1954classification}
Petrov, A.~Z. 1954, Uchenye Zapiski Kazanskogo Universiteta. Seriya
  Fiziko-Matematicheskie Nauki, 114, 55

\bibitem[{{Phinney}(1984)}]{1984PhDT........92P}
{Phinney}, III, E.~S. 1984, PhD thesis, University of Cambridge, UK

\bibitem[{Qiu {et~al.}(2023)Qiu, Ricarte, Narayan, Wong, Chael, \&
  Palumbo}]{Qiu:2022fzl}
Qiu, R., Ricarte, A., Narayan, R., {et~al.} 2023, Mon. Not. Roy. Astron. Soc.,
  520, 4867, \dodoi{10.1093/mnras/stad466}

\bibitem[{Ricarte {et~al.}(2022)Ricarte, Palumbo, Narayan, Roelofs, \&
  Emami}]{Ricarte:2022wpd}
Ricarte, A., Palumbo, D. C.~M., Narayan, R., Roelofs, F., \& Emami, R. 2022,
  Astrophys. J. Lett., 941, L12, \dodoi{10.3847/2041-8213/aca087}

\bibitem[{Ricarte {et~al.}(2020)Ricarte, Prather, Wong, Narayan, Gammie, \&
  Johnson}]{Ricarte:2020llx}
Ricarte, A., Prather, B.~S., Wong, G.~N., {et~al.} 2020, Mon. Not. Roy. Astron.
  Soc., 498, 5468, \dodoi{10.1093/mnras/staa2692}

\bibitem[{Rybicki \& Lightman(1979)}]{1979Lightman}
Rybicki, G.~B., \& Lightman, A.~P. 1979, Lightman Radiative Processes in
  Astrophysics (Lightman Radiative Processes in Astrophysics)

\bibitem[{Tchekhovskoy {et~al.}(2010)Tchekhovskoy, Narayan, \&
  McKinney}]{Tchekhovskoy:2009ba}
Tchekhovskoy, A., Narayan, R., \& McKinney, J.~C. 2010, Astrophys. J., 711, 50,
  \dodoi{10.1088/0004-637X/711/1/50}

\bibitem[{Tchekhovskoy {et~al.}(2011)Tchekhovskoy, Narayan, \&
  McKinney}]{Tchekhovskoy:2011zx}
---. 2011, Mon. Not. Roy. Astron. Soc., 418, L79,
  \dodoi{10.1111/j.1745-3933.2011.01147.x}

\bibitem[{Thorne \& Macdonald(1982)}]{thorne1982electrodynamics}
Thorne, K.~S., \& Macdonald, D. 1982, Monthly Notices of the Royal Astronomical
  Society, 198, 339

\bibitem[{Vincent {et~al.}(2021)Vincent, Wielgus, Abramowicz, Gourgoulhon,
  Lasota, Paumard, \& Perrin}]{Vincent:2020dij}
Vincent, F.~H., Wielgus, M., Abramowicz, M.~A., {et~al.} 2021, Astron.
  Astrophys., 646, A37, \dodoi{10.1051/0004-6361/202037787}

\bibitem[{Walker \& Penrose(1970)}]{Walker:1970un}
Walker, M., \& Penrose, R. 1970, Commun. Math. Phys., 18, 265,
  \dodoi{10.1007/BF01649445}

\bibitem[{Zhang {et~al.}(2024)Zhang, Ricarte, Pesce, Johnson, Nagar, Narayan,
  Ramakrishnan, Doeleman, \& Palumbo}]{Zhang:2024owe}
Zhang, X.~A., Ricarte, A., Pesce, D.~W., {et~al.} 2024.
\newblock \doarXiv{2406.17754}

\end{thebibliography}
\bibliographystyle{aasjournal}

\end{document}